%% file: fwd_PV.tex
\documentclass[fleqn,usenatbib]{mnras}

\usepackage{newtxtext,newtxmath}

\usepackage[T1]{fontenc}

\usepackage{xspace}

\DeclareRobustCommand{\VAN}[3]{#2}
\let\VANthebibliography\thebibliography
\def\thebibliography{\DeclareRobustCommand{\VAN}[3]{##3}\VANthebibliography}

\usepackage{graphicx}	
\usepackage{amsmath}	
\usepackage{bm}

\newcommand{\diffop}{\mathrm{d}}
\newcommand{\mmat}[1]{{\mathbf{#1}}}
\newcommand{\kms}{km~s$^{-1}$}
\newcommand{\mD}{\mathcal{D}}
\newcommand{\mP}{\mathcal{P}}
\newcommand{\mM}{\mathcal{M}}
\newcommand{\mN}{\mathcal{N}}

\newcommand{\borg}{{\sc borg}\xspace}
\newcommand{\virbius}{{\sc virbius}}
\newcommand{\mvec}[1]{{\bm{#1}}}
\newcommand{\tbulk}{\text{bulk}}

\title[Bayesian peculiar velocity reconstruction]{Reconstructing dark matter distribution with peculiar velocities: Bayesian forward modelling with corrections for inhomogeneous Malmquist bias}

\author[S. S. Boruah, G. Lavaux and M. J. Hudson]{Supranta S. Boruah$^{1}$\thanks{Contact e-mail: \href{mailto:ssarmabo@arizona.edu}{ssarmabo@arizona.edu}}, Guilhem Lavaux$^{2}$\thanks{Contact e-mail: \href{mailto:guilhem.lavaux@iap.fr}{guilhem.lavaux@iap.fr}}
Michael J. Hudson$^{3,4,5}$\thanks{Contact e-mail: \href{mailto:mike.hudson@uwaterloo.ca}{mike.hudson@uwaterloo.ca}}
\\
$^{1}$ Department of Astronomy and Steward Observatory, University of Arizona, 933 N Cherry Ave, Tucson, AZ 85719, USA \\
$^{2}$CNRS \& Sorbonne Universit\'e, UMR7095, Institut d'Astrophysique de Paris, F-75014, Paris, France \\
$^{3}$Waterloo Centre for Astrophysics, University of Waterloo,  200, University Ave W, Waterloo, ON N2L 3G1\\
$^{4}$Department of Physics and Astronomy, University of Waterloo, Waterloo, ON, N2L 3G1, Canada\\
$^{5}$Perimeter Institute for Theoretical Physics, 31 Caroline St N, Waterloo, ON N2L 2Y5
}

\date{Accepted XXX. Received YYY; in original form ZZZ}

\pubyear{2021}

\begin{document}
\label{firstpage}
\pagerange{\pageref{firstpage}--\pageref{lastpage}}
\maketitle

\begin{abstract}
We present a forward-modelled velocity field reconstruction algorithm that performs the reconstruction of the mass density field using only peculiar velocity data. Our method consistently accounts for the inhomogeneous Malmquist bias using analytic integration along the line-of-sight. By testing our method on a simulation, we show that our method gives an unbiased reconstruction of the velocity field. We show that not accounting for the inhomogeneous Malmquist bias can lead to significant biases in the forward-modelled reconstructions. We applied our method to a peculiar velocity data set consisting of the SFI++ and 2MTF Tully-Fisher catalogues and the A2 supernovae compilation, thus obtaining a novel velocity reconstruction in the local Universe. Our velocity reconstructions have a cosmological power spectrum consistent with the theoretical expectation. Furthermore, we obtain a full description of the uncertainties on reconstruction through samples of the posterior distribution. We validate our velocity reconstruction of the local Universe by comparing it to an independent reconstruction using the 2M++ galaxy catalogue, obtaining good agreement between the two reconstructions. Using Bayesian model comparison, we find that our velocity model performs better than the adaptive kernel smoothed velocity with the same peculiar velocity data. However, our velocity model does not perform as well as the velocity reconstruction from the 2M++ galaxy catalogue, due to the sparse and noisy nature of the peculiar velocity tracer samples. The method presented here provides a way to include peculiar velocity data in initial condition reconstruction frameworks.
\end{abstract}



\input{sections/intro}
\input{sections/data}
\input{sections/method}
\input{sections/sim_validation}
\input{sections/real_data}
\input{sections/discussion}
\input{sections/summary}
\section*{Acknowledgement}
We thank Eleni Tsaprazi, Florent Leclercq and James Prideaux-Ghee for useful feedback on the manuscript. This work has been done as part of the activities of the Domaine d'Int\'{e}r\^{e}t Majeur (DIM) ``Astrophysique et Conditions d'Apparition de la Vie'' (ACAV), and received financial support from R\'{e}gion Ile-de-France. GL acknowledges financial support from the ANR BIG4, under reference ANR-16-CE23-0002.  MH acknowledges the support of an NSERC Discovery Grant. This work is done within the Aquila Consortium\footnote{\url{https://www.aquila-consortium.org/}}.
This work has made use of the Infinity Cluster hosted by Institut d'Astrophysique de Paris. We thank Stephane Rouberol for running this cluster smoothly for us.

\section*{Data availability statement}
The Second Amendment (A2) supernovae compilation is available with the supplementary data of \url{https://doi.org/10.1093/mnras/staa2485}. The 2MTF and the SFI++ catalogues are publicly available with their respective publications that are cited in section \ref{sec:data}. The 2M++ reconstruction used in this work is publicly available at \url{https://cosmicflows.iap.fr/}. The data products generated in this work will be shared on reasonable request to the authors.


\bibliographystyle{mnras}
\bibliography{fwd_PV} 
\appendix
\input{sections/appendix/velocity_error}


\bsp	
\label{lastpage}
\end{document}

%% file: sections/intro.tex
\section{Introduction}\label{sec:intro}
 The large-scale structure of the Universe sources the peculiar velocity of the galaxies. The study of the peculiar velocity of galaxies in the local Universe is important in cosmological applications for two reasons: {\it i)} peculiar velocities are the only probe of the growth of large-scale structure in the low-redshift ($z \lesssim 0.1$) Universe, {\it ii)} peculiar velocities are a nuisance parameter in the measurement of the Hubble parameter $H_0$ and thus need to be corrected for in such measurements. Correcting for the peculiar velocity contributions to the redshifts in the measurement of the expansion history relies on using a reconstruction of the peculiar velocity field in the local Universe. Given the current tension in the measurement of the Hubble constant, $H_0$ \citep{Verde2019}, we need to account for all sources of uncertainties. In order to obtain unbiased estimates of the peculiar velocity corrections for the local $H_0$ measurement, we need accurate models of the local peculiar velocity field.

Therefore, reconstruction of the velocity field of the local Universe is important for these cosmological applications. Many methods of reconstruction of the peculiar velocity field have been proposed in the literature. We can categorize the different methods according to two criteria: {\it i)} the data used for the reconstruction, i.e, whether galaxy catalogues or peculiar velocity tracers are used in the reconstruction and {\it ii)} whether the method involves forward-modelling or a direct inversion from the data.

 Most commonly used reconstructions of the local velocity field uses galaxy catalogues to reconstruct the galaxy density field \citep[e.g, ][]{Carrick_et_al, Lilow2021}. Following the reconstruction of the galaxy density field, the velocity field is calculated using linear perturbation theory. However, since galaxies are biased tracers of the matter field, we need to fit for the parameter $\beta = f/b$ to get the velocity field, where $f$ is the logarithmic growth parameter and $b$ is the linear galaxy bias. This is done by comparing the reconstructed velocity field to the peculiar velocity data. Galaxy catalogues are much denser than peculiar velocity catalogues. Furthermore, the velocity estimates obtained from peculiar velocity tracers are very noisy. Therefore, reconstruction of the density field is usually an easier task than the direct reconstruction of the velocity field. However, connecting the galaxy density to the total matter density requires us to assume a model of the galaxy bias, which is an important source of systematic uncertainty in these reconstruction. On the other hand, using the peculiar velocity tracer directly for reconstruction does not require that assumption, thus providing a complementary method for reconstruction of the velocity field.  One well-known example is the POTENT method \citep{POTENT, Bertschinger1990} which uses peculiar velocity data directly to reconstruct the velocity field assuming that the velocity field is the gradient of the gravitational potential field. Another such widely used method is the adaptive kernel-smoothed velocity reconstruction method \citep{Springob2014, Springob2016} that smooths the peculiar velocity data to obtain the velocity reconstruction. 
 
 The other criteria that we can use to categorize the reconstruction methods is whether the reconstruction is performed using forward-modelling or uses a direct inversion from the data. Inverting non-linear problems from partial, noisy, observations is an ill-posed inverse problem, which makes forward-modelled methods particularly suitable for the task of reconstruction of high-dimensional fields. Forward-modelled reconstruction methods have become increasingly popular in cosmology and have been applied in a range of different applications such as initial conditions reconstruction \citep{Jasche2013, Modi2018, Jasche2019}, weak lensing \citep{Porqueres2021a, Porqueres2021b, Fiedorowicz2021} and CMB lensing \citep{Millea2020b, Millea2020a}. Such methods have also been used for the local velocity field reconstruction. The simplest of such methods uses a Wiener filtering technique \citep{Zaroubi1999}. This approach assumes that the density/velocity field is described as a Gaussian random field and the Wiener filtered reconstruction is the maximum-a-posteriori (MAP) solution for the problem. The Wiener filtering approach has been extended to account for uncertainties using a constrained realization approach \citep{HoffmanRibak1991, HoffmanCourtoisTully2015, HoffmanCarlesiPomarede2018, Lilow2021} and to account for biases in Wiener filtering using the unbiased minimal variance approach \citep{Zaroubi2002}. Another similar approach is the Bayesian hierarchical method, \virbius ~\citep{Lavaux2016} which is based on the constrained realization approach but accounts for many different systematic effects in its analysis. This approach has been been applied to the \textit{Cosmicflows-3} \citep{Tully2016} dataset by \citet{Graziani2019}. However, this method fails to account for the inhomogeneous Malmquist bias (IHM) which is an important source of systematic error in peculiar velocity analysis. The IHM bias arises from an incorrect assumption on the distribution of peculiar velocity tracers due to neglecting the line-of-sight inhomogeneities. In this paper, we introduce a forward-modelled reconstruction method that uses peculiar velocity data in its reconstruction while consistently accounting for the inhomogeneous Malmquist bias in the analysis. As we show in this paper, not accounting for the IHM bias in the reconstruction may lead to substantial biases in the reconstruction. One approach to correct for the IHM bias is by marginalizing over an accurate model of galaxy distribution \citep[e.g,][]{Hudson1994}. In this work, we use such an approach to deal with the inhomogeneous Malmquist bias within our forward-modelled peculiar velocity reconstruction method. Our reconstruction results in samples of reconstruction that are compatible with the data from a given peculiar velocity catalogue, and sampled from a posterior assuming a $\Lambda$CDM Gaussian prior. Furthermore, the full correlated uncertainties in the reconstruction can be estimated from the posterior samples. Our method presented here can be extended in a straightforward manner to include more complex gravity models such as initial condition reconstruction methods like \borg \citep{Jasche2013, Jasche2019}.

The paper is structured as follows: in Section \ref{sec:data}, we describe the simulations and data used in this paper. In Section \ref{sec:method}, we describe our reconstruction method in detail. Our method is validated with mock simulated catalogues in Section \ref{sec:sim_results} followed by an application of our method to reconstruct the velocity field of the local Universe in Section \ref{sec:vel_rec}. Finally, after a brief discussion in Section \ref{sec:discussion}, we summarize our results in Section \ref{sec:summary}. In Appendix \ref{app:error_origin}, we investigate in further detail, the various sources contributing to the reconstruction error.

%% file: sections/data.tex
\section{Simulations and data}\label{sec:data}

In this section, we describe the simulations and the peculiar velocity surveys used in this work. 

\subsection{VELMASS simulation}\label{ssec:velmass}
In order to validate our method, we use an $N$-body simulation  from the VELMASS suite of simulation.\footnote{For more details on the simulation which we used, see \citet{halo_painting}} The simulation was performed in a cubic box of size $(2~h^{-1}$ Gpc)$^3$ with $2048^3$ particles with mass $9.387\times 10^{10}~h^{-1}M_{\odot}$. The cosmological parameters used are: $\Omega_m= 0.315, \Omega_b = 0.049, H_0 = 68$ km s$^{-1}$ Mpc$^{-1}$, $\sigma_8 = 0.81$, $n_s = 0.97$ and $Y_{\text{He}} = 0.248$. The {\sc rockstar} \citep{rockstar} halo finding software was used to identify the dark matter halos in the simulation. We only consider the halos with mass $M > 3\times10^{12}~h^{-1} \mathrm{M}_{\odot}$.

In order to create a mock peculiar velocity survey, we populate the halos with mass $M > 3\times10^{12}~h^{-1}\mathrm{M}_{\odot}$ with a standard candle, with different levels of intrinsic scatter. That is, we assign it an absolute magnitude $\mM=\mM_0 + \epsilon$, where $\mM_0$ is the standard candle absolute magnitude and $\epsilon$ is the intrinsic scatter drawn from a normal distribution, $\epsilon \sim \mN(0, \sigma^2_{\text{int}})$. Here, $\sigma_{\text{int}}$ is the intrinsic scatter. We can calculate the apparent magnitude from this absolute magnitude and the distance to the halos. We make mock catalogues with different levels of intrinsic scatter. Finally, we make an apparent magnitude cut to create a magnitude limited catalogue. From this magnitude limited sample, we randomly select $7\,000$ halos. While running the forward likelihood method described in section \ref{ssec:prob_model}, the method requires the estimated galaxy density field as an input. For this purpose, we construct the halo density field and smooth it using a Gaussian filter of smoothing length $4~h^{-1}$ Mpc.

\subsection{Peculiar velocity surveys}\label{ssec:pv_surveys}

\begin{figure}
    \centering
    \includegraphics[width=\linewidth]{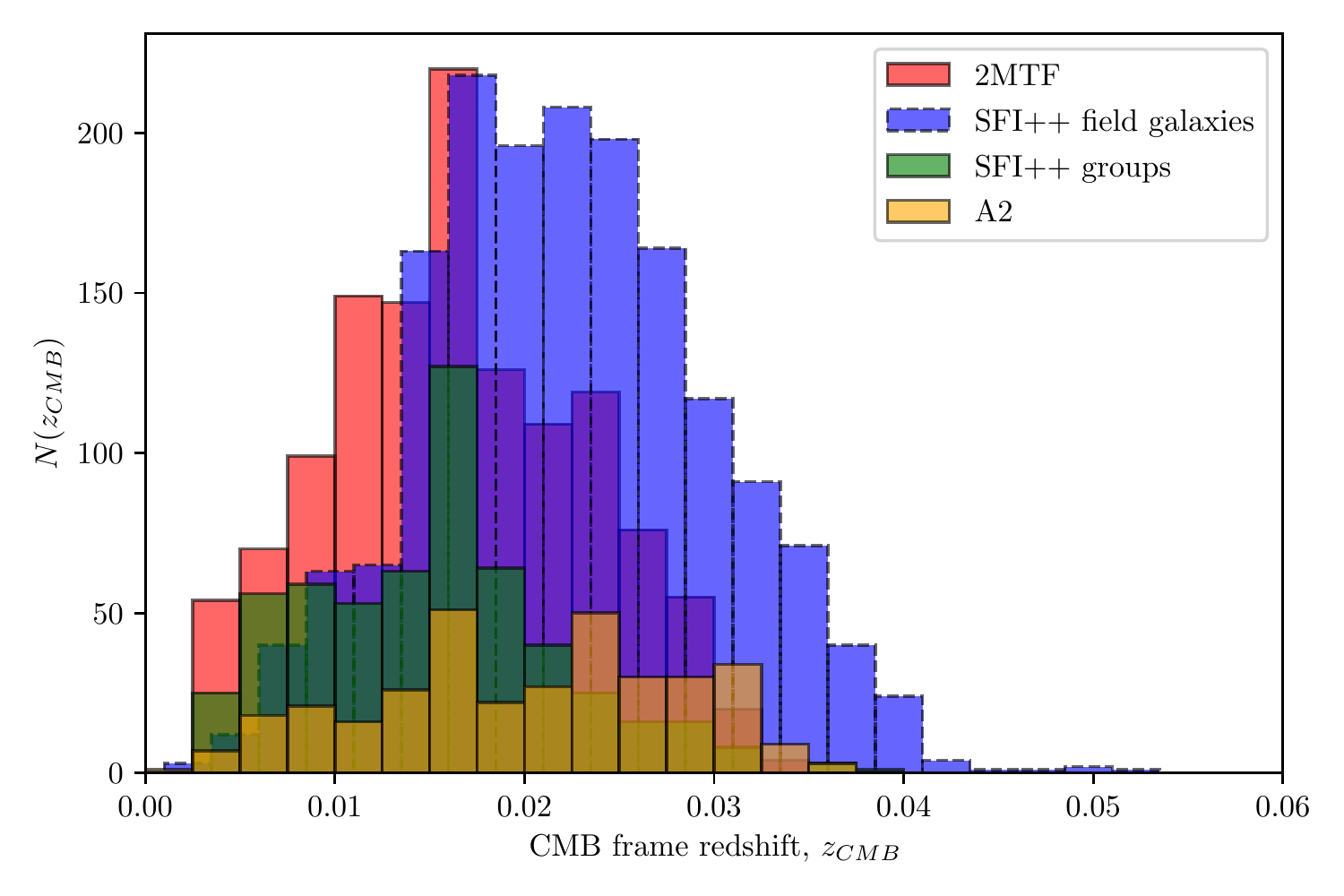}
    \caption{Redshift distribution of the tracers in the different peculiar velocity catalogues used in this work. We have imposed a distance cut of $d < 100~h^{-1}$ Mpc on all catalogues.}
    \label{fig:z_dist}
\end{figure}

In this work, we use peculiar velocity data from various surveys in our algorithm to reconstruct the velocity field of the local Universe. The three peculiar velocity data sets that we use in this work are the A2 supernovae compilation, SFI++ and 2MTF Tully-Fisher (TF) catalogues. Compared to  our treatment of these data-sets in \citet{Boruah2020}, we do not fit the TF or the SNe light curve fitting parameters during the reconstruction. Instead we use the distances as provided in the catalogues. We however fit for a scaling factor that may arise due to miscalibration in the fitting of the zero-points of the TF relations as detailed in section \ref{ssec:prob_model}. Furthermore, in order to limit the edge effects in the reconstruction, we impose a distance cut of $d < 100~h^{-1}$ Mpc. Note that imposing such a distance cut does not introduce a selection bias in the forward peculiar velocity analysis methods \citep{StraussWillick95}. The redshift distribution of peculiar velocity tracers is shown in Figure ~\ref{fig:z_dist}. In the following, we provide a brief description of the data used in this work. We refer readers to \citet{Boruah2020, Boruah2021} for more details on the selection and the treatment of outliers, choosing only to highlight some main features of the data sets. 

\begin{itemize}
\item \textit{\underline{A2 Supernovae}}:
In \cite{Boruah2020}, we compiled the second amendment (A2) dataset of nearby supernovae from publicly available supernovae from the CfA supernovae sample \citep{Hicken2009}, Carnegie Supernovae Project - Data Release 3 \citep[CSP-DR3,][]{Krisciunas2017}, the Lick observatory Supernova Survey \citep[LOSS, ][]{Ganeshalingam2013} and the Foundation sample \citep{Foley2018, Jones2019} of supernovae. We only consider the supernovae with estimated distance, $d < 100~h^{-1}$ Mpc.

\item \textit{\underline{SFI++}}:
We use peculiar velocity data from the SFI++ \citep{Masters2006, Springob2007} Tully-Fisher catalogue, which is an $I$-band TF survey. We follow the treatment of \citet{Boruah2020} to remove outliers. For galaxies in groups, we use the peculiar velocity data for the groups to suppress non-linear contributions to the peculiar velocity data. We only consider the galaxies/groups with estimated distance, $d < 100~h^{-1}$ Mpc.
\item \textit{\underline{2MTF}}:
The 2MTF survey \citep{Masters2008, Hong2019} is an all sky Tully-Fisher in the near-infrared $J, H$ and $K$ bands. The survey is limited to $cz < 10000$ km/s.
\end{itemize}

With the given distance cuts, we select $345$,  $1682$, $556$, $1225$ objects from the A2 catalogue, SFI++ field galaxy catalogue, SFI++ group catalogue and 2MTF catalogue respectively.

%% file: sections/method.tex
\section{Methodology}\label{sec:method}

We introduce a novel method for forward modelled velocity reconstruction which can consistently account for inhomogeneous Malmquist (IHM) bias in the reconstruction in this paper. In this section, we describe the theory and the methodology behind our method for reconstructing the velocity field with the peculiar velocity data. 

\subsection{Peculiar velocity theory}

\begin{figure}
    \centering
    \includegraphics[width=0.6\linewidth]{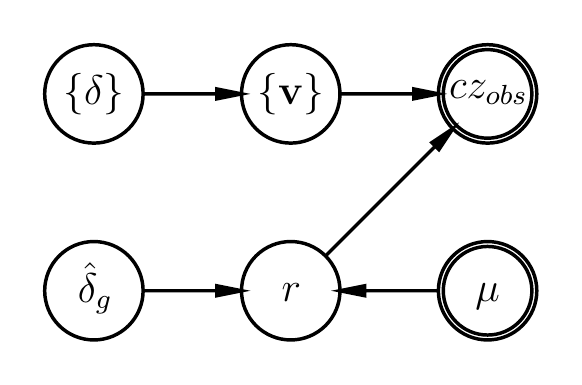}
    \caption{A probabilistic graph showing the dependency of the different observables and the underlying variables. The observables (redshift, $cz_{\text{obs}}$ and the distance modulus, $\mu$) are shown with a double circle. In the forward  methods of peculiar velocity analysis (see section \ref{ssec:prob_model}), the redshift of the peculiar velocity tracers are predicted as a function of the distance estimates. This is reflected in the relationship between the observables. Note that, we use an external estimate of the galaxy density field, $\hat{\delta}_g$ in order to correct for the inhomogeneous Malmquist bias.}
    \label{fig:pgm}
\end{figure}

In the $\Lambda$CDM model, under linear perturbation theory, the present day peculiar velocity field can be expressed in terms of the density field as \citep{Peebles1980}, 
\begin{equation}\label{eqn:linear_theory_v}
    \mvec{v}_{\mvec{k}} = \frac{if H_0 \mvec{k}}{k^2}\delta_{\mvec{k}},
\end{equation}
where $f\approx\Omega_m^{0.55}$ is the logarithmic growth rate in the $\Lambda$CDM model, $i^2=-1$. Since a factor of $k$ appears in the denominator, the peculiar velocity field is sensitive to the large-scale density modes. Since these modes are well described with linear perturbation theory, linear theory predictions of peculiar velocity provide a fairly accurate result for the true velocity field. The peculiar velocity estimated using linear theory has been calibrated with simulations \citep{Carrick_et_al, HollingerHudson2021} and the uncertainty in the linear theory velocity estimates due to the non-linearities was found to be $\sigma_{\text{NL}} \approx 150$~\kms. In this work, we use linear perturbation theory to predict the velocity field. In the future, this can be extended to include non-linear differentiable velocity field models such as the Lagrangian perturbation theory \citep[LPT,][]{Bouchet94} full particle mesh or COLA \citep{Tassev2013} simulations that have previously been used for forward-modelled density reconstruction \citep[e.g, ][]{Jasche2019}. 

\subsection{Inhomogeneous Malmquist bias}\label{ssec:ihm_bias}

Peculiar velocity analyses are impacted by a number of different statistical biases, including the homogeneous and inhomogeneous Malmquist biases. The homogeneous Malmquist bias arises due to the fact that the number of galaxies grows along the line-of-sight according to the volume factor. In this paper, we mainly discuss the impact of inhomogeneous Malmquist bias in forward modelled reconstruction with peculiar velocity data. Therefore, we discuss the basics of inhomogeneous Malmquist (IHM) bias in some detail in this section.

The IHM bias arises due to the lack of accounting for inhomogeneity in the radial distribution of the peculiar velocity tracers. Neglecting the overdensities along the line-of-sight leads to a bias in the inferred distance, $d_{\text{est}}$ compared to the true distance, $d_{\text{true}}$. The magnitude of the bias is given as \citep{LyndenBell1988, StraussWillick95}, 
\begin{equation}\label{eqn:ihm_bias}
    \Delta_{\text{IHM}} = \langle d_{\text{true}}  \rangle - \langle d_{\text{est}} \rangle \approx d_{\text{est}} \gamma \Delta_d^2,
\end{equation}
where, $ \Delta_d$ is the fractional uncertainty in the distance estimates and $\gamma$ is the logarithmic slope in the galaxy over-density along the line-of-sight given by, 
\begin{equation}
    \gamma = \frac{\diffop \ln n(r)}{\diffop \ln r}.
\end{equation}
Since the slope changes sign in front of and behind an overdensity, the distance of the galaxies in front of the overdensity is underestimated and those at the back are overestimated. This may be interpreted as a spurious flow if the IHM bias correction is neglected and thus biases the magnitude of the velocity field to a higher value compared to the true value. Therefore, the inferred value of $\beta$ is biased high when the IHM bias is not accounted for in velocity-velocity comparison methods \citep[e.g.][]{Boruah2020}. As we will show in section \ref{sec:sim_results}, in forward-modelled reconstruction from peculiar velocity data, this effect shows up as an increment in the inferred power spectrum of the reconstructed density field if the IHM bias is not corrected for.

\subsection{Probability model}\label{ssec:prob_model}

The method we present in this paper is a forward-modelled Bayesian method to reconstruct the velocity field using the data from a peculiar velocity survey. One such reconstruction method is the {\sc virbius} algorithm \citep{Lavaux2016} which has previously been applied to the \textit{Cosmicflows-3} \citep{Tully2016} dataset in \citet{Graziani2019}. In their method, the true distance of the peculiar velocity tracers are used as latent parameters and the value of the distance is inferred in each Markov Chain Monte Carlo (MCMC) step. The homogeneous Malmquist bias is dealt with by using a skewed prior on these distances. However, the inhomogeneous Malmquist bias is not explicitly accounted for in their analysis. In this work, we account for the inhomogeneous Malmquist bias by using analytic integration over the line-of-sight uncertainty. Marginalizing over the true distance also reduces the complexity since, we do not have to infer their value in each MCMC steps. Furthermore, the posterior distribution of the peculiar velocity tracer distances can potentially be highly multi-modal due to the line of sight inhomogeneities, which makes the analytical marginalization preferable to sampling based methods. 

In the forward modelled reconstruction, we want to infer the velocity field, $\{\mvec{v}\}$ given the peculiar velocity data. The peculiar velocity data set consists of the observed redshift, $\{z_{\text{obs}}\}$, and the estimate of its distance modulus, $\{\mu\}$, for a set of peculiar velocity tracers. The uncertainty associated with the distance modulus is denoted by $\sigma_{\mu}$. In linear perturbation theory, the velocity field is determined in terms of the density field, $\{\delta\}$ through     equation \eqref{eqn:linear_theory_v}. Therefore, we can equivalently think about the process described here as a density reconstruction algorithm. Furthermore, we need an estimate of the galaxy density field to correct for the inhomogeneous Malmquist bias. For this work, we provide the estimate of the galaxy density field, $\{\hat{\delta}_g\}$, as an input to our method.

\begin{figure*}
    \centering
    \includegraphics[width=0.88\linewidth]{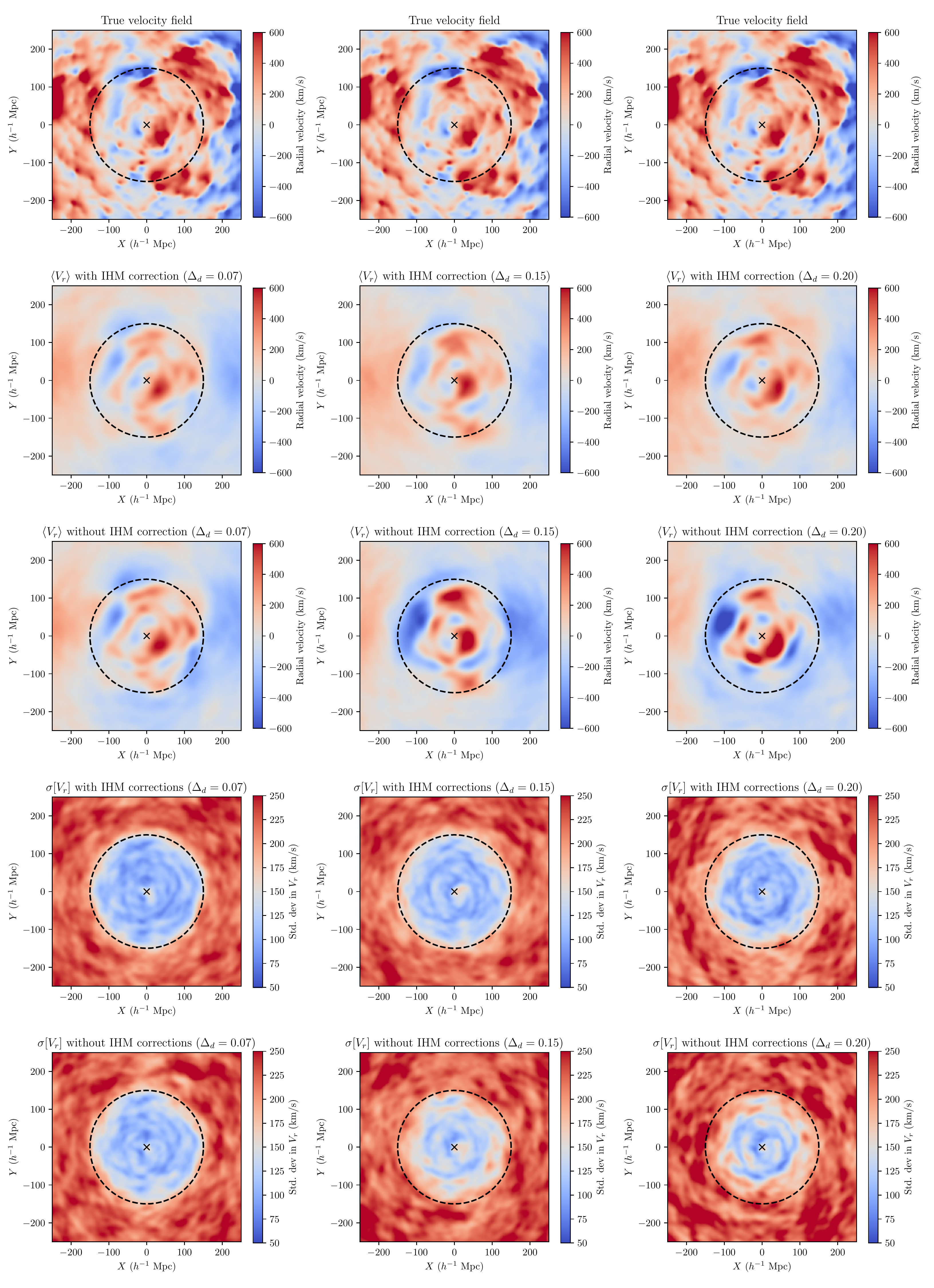}
    \caption{A visual comparison of the reconstructed radial velocity field and the associated uncertainties with the true velocity field in the $Z=0$ plane in our mock simulation. We show the reconstruction for $3$ different mock surveys with a distance error of $\Delta_d=0.07$ ({\it left}), $\Delta_d=0.15$ ({\it centre}) and $\Delta_d=0.20$ ({\it right}). 
    The top row is the true velocity field in the simulation in the $Z=0$ plane. The black dashed circle is at a radius of $150~h^{-1}$ Mpc, which is close to the $95\%$ completeness radius for each survey scenario. The second and the third rows show the mean velocity for the samples of our reconstructions with and without inhomogeneous Malmquist bias corrections respectively. 
    As can be seen, if we do not correct for the IHM bias, the velocity field in the data region is boosted up compared to the true velocity field. The bottom two rows shows the standard deviation in the reconstruction samples with and without the IHM bias correction.}
    \label{fig:visual_comparison}
\end{figure*}

In this work, we use a variant of the forward VELMOD method \citep{velmod}, which was introduced in \citet{PikeHudson2005} to model the likelihood. For simplicity, we call this likelihood the \textit{Forward Likelihood}. The forward likelihood method corrects for the inhomogeneous Malmquist bias by accounting for the density dependent line-of-sight distribution of peculiar velocity tracers. In the forward likelihood method, the observed redshift is predicted as a function of the distance estimates. That is, we model the likelihood using the conditional probability, $\mP(\{cz_{\text{obs}}\}|\{\mu\}, \{ \mvec{v} \}, \{ \hat{\delta}_g\})$. We show the dependency of the various variables and the observables in a probabilistic graph in Figure~\ref{fig:pgm}. Using Bayes theorem, the posterior for the density field $\delta$ can be written as 
\begin{equation}\label{eqn:posterior}
    \mP(\{\delta\}|\{cz_{\text{obs}}\}, \{\mu\}, \{\hat{\delta}_g\}) \propto \mP(\{cz_{\text{obs}}\}|\{\mu\}, \{\delta\}, \{\hat{\delta}_g\}) \mP(\{\delta\}).
\end{equation}
In our runs, we use a $128^3$ cubic box where the density is inferred. That leads to a total of $128^3$ density parameters. In order to sample from such high-dimensional parameter space, we used the Hamiltonian Monte-Carlo \citep[HMC, ][]{HMC1, HMC2} algorithm. HMC can be used to sample from very high dimensional parameter space by using the gradient information of the log-posterior. In this paper, we implement our code by relying on the {\sc jax} package \citep{jax} which includes the gradients of the requisite function by using automatic differentiation. Sampling with HMC is highly sensitive to the choice of mass matrix. The optimal choice of mass matrix is to use inverse of the width of the posterior as the mass matrix \citep{Taylor2008}. We use automatic differentiation to calculate the second derivative of the posterior for a number of Gaussian mocks and use it to calculate the mass matrix. 

The RHS of equation \eqref{eqn:posterior} consists of two terms, {\it i)} the prior on the density field, $\mP(\{\delta\})$ and {\it ii)} The likelihood, $\mP(\{cz_{\text{obs}}\}|\{\mu\},\{\delta\}, \{\hat{\delta}_g\})$. We will first discuss the prior we use in this work. In this work, we assume a Gaussian prior on the density field. In the Fourier space, we can write it as
\begin{equation}\label{eqn:density_prior}
    \mP(\{\delta\}) = \prod_{\mvec{k}}\frac{1}{\sqrt{2\pi \sigma^2_{k}}}\exp\bigg[ -\frac{|\delta_{\mvec{k}}|^2}{2 \sigma^2_{k}}\bigg],
\end{equation}
where $\sigma^2_k=V_s P(k)$ is the variance in $\delta_{\mvec{k}}$. Here, $P(k)$ is the $\Lambda$CDM power spectrum and $V_s$ is the volume of the simulation box. In this work, we fix our cosmological parameters to the fiducial cosmological parameters used in the VELMASS simulation. We note that the present day density field is significantly non-Gaussian. Thus it cannot be well-described using a Gaussian prior. In the future, we can incorporate the method presented here with the {\sc borg} framework which uses a non-linear, differentiable gravity model to forward model the present day velocity field from the density field at an initial time when the field is well-described as a Gaussian field. Nevertheless, due to the sensitivity of the velocity field to the large-scale density modes, the Gaussian prior still proves to be a useful approximation.

Assuming that the random components of the radial velocity for the various peculiar velocity tracers are independent, we can write,
\begin{equation}
    \mP(\{cz_{\text{obs}}\}|\{\mu\},\{\delta\}, \{\hat{\delta}_g\}) = \prod_{i=1}^{N_{\text{PV}}} \mP(cz_{\text{obs}}^{(i)}|\mu^{(i)},\{\delta\}, \{\hat{\delta}_g\})
\end{equation}
 Given the large uncertainties on the distance estimates, we marginalize over the true line-of-sight distance, $r$, for the peculiar velocity tracer. The likelihood for the individual peculiar velocity tracers can be expressed as
\begin{align}\label{eqn:fwd_lkl}
    \mP(cz^{(i)}_{\text{obs}}| \mu^{(i)},&\{\delta\}, \{\hat{\delta}_g\}) \nonumber \\ 
    &= \int_0^{\infty}\diffop r \mP(cz^{(i)}_{\text{obs}}|r, \{\mvec{v}\})\mP(r|\mu^{(i)}, \{\hat{\delta}_g\}).
\end{align}
In the above, $ \{\mvec{v}\}$ is the linear velocity field calculated from the density field $\{ \delta \}$ using equation \eqref{eqn:linear_theory_v}.
The first term inside the integral is modelled as a Gaussian with the uncertainty in the observed redshift given by $\sigma_{\text{NL}}$, which is the uncertainty in the velocity induced by the non-linearities in the density field. The likelihood function in the integrand is given by
\begin{equation}
    \mP(cz^{(i)}_{\text{obs}}|r, \{\mvec{v}\}) = \frac{1}{\sqrt{2\pi \sigma^2_{\text{NL}}}}\exp \bigg[ -\frac{(cz^{(i)}_{\text{obs}} - cz^{(i)}_{\text{pred}}(r, \mvec{v}))^2}{2\sigma^2_{\text{NL}}}\bigg],
\end{equation}
where
\begin{equation}
    1 + z_{\text{pred}}(r, \mvec{v}) = [1 + z_{\text{cos}}(r)]\bigg(1 + \frac{v_r(r)}{c}\bigg).
\end{equation}
In the above equation, $z_{\text{cos}}$ denotes the `cosmological' recessional redshift. In order to get the estimate of the velocity field, we smooth the linear velocity field using a Gaussian smoothing filter with a smoothing scale of $R_{\text{smooth}} = 4 h^{-1}$ Mpc. We use this smoothing scale for all subsequently reported velocity fields. The velocity field smoothed at this scale is found to be unbiased in $N$-body simulations and has a velocity uncertainty of $\sigma_{\text{NL}} \approx 150$ km/s \citep{Carrick_et_al, HollingerHudson2021}. We fix the value of $\sigma_{\text{NL}}$ to this value in our inference.

The `forward' method for peculiar velocity analysis is susceptible to homogeneous and inhomogeneous Malmquist biases \citep{Hudson1994, StraussWillick95}. As mentioned earlier, the homogeneous Malmquist bias arises if we do not account for the fact that there are more galaxies at large radius. Moreover, galaxies are not uniformly distributed in the Universe, instead are clustered at regions of large overdensities. Neglecting these overdensities along the line-of-sight leads to the inhomogeneous Malmquist bias as discussed in section \ref{ssec:ihm_bias}. We use an estimate of the line-of-sight galaxy density, $\hat{\delta}_g(r)$, to account for the inhomogeneous Malmquist bias in the radial distribution. Using the galaxy density field, the expected number of peculiar velocity tracers along the line of sight can be estimated as
\begin{equation}
    n(r| \{\hat{\delta}_g\}) \propto r^2 [1 + \hat{\delta}_g(r)].
\end{equation}
Accounting for this dependence on the line-of-sight density, we can write the distribution, 
\begin{equation}\label{eqn:los_distribution}
    \mP(r|\mu^{(i)}, \{\hat{\delta}_g\}) \propto n(r| \{\hat{\delta}_g\})  \exp\bigg(-\frac{[\mu^{(i)} - \mu(r)]^2}{2\sigma^{(i)2}_{\mu}} \bigg),
\end{equation}
where $\mu(r) = 5\log_{10}(r/10~\text{pc})$ and $\sigma_{\mu}^{(i)}$ is the uncertainty on the estimate of the distance modulus. Note that we assume a Gaussian uncertainty on the distance modulus. One of the main sources of systematic uncertainty in estimating the distances to the peculiar velocity tracers is the calibration of the zero-point of the distance-luminosity relation. In order to take care of this uncertainty, we introduce a scale factor, $A$, so that all the distance estimates in a given catalog is scaled by the same factor, $\mu \rightarrow \mu + 5\log_{10}A$ (or equivalently $d \rightarrow A d$). We then fit and marginalize over this factor during the reconstruction in a block sampling scheme. When using multiple catalogues, we fit a separate scaling factor for different catalogues.

In this work, the estimated galaxy density field, $\{\hat{\delta}_g\}$ is an additional input to our reconstruction code and is assumed to be independent of the density field that is inferred. This is not fully self-consistent and needs to be improved in the future. We note that we do not use any cross-correlation information between the estimated galaxy density field and the inferred velocity field. The estimated galaxy density field simply provides an estimate for the radial distance of tracer. For the test with simulation, we use the true underlying particle density to account for the IHM correction. For our run with the real data, we use an iteratively reconstructed density field from \citet{Carrick_et_al}. 

%% file: sections/sim_validation.tex
\section{Validation with  simulations}\label{sec:sim_results}

\begin{figure*}
    \centering
    \includegraphics[width=\linewidth]{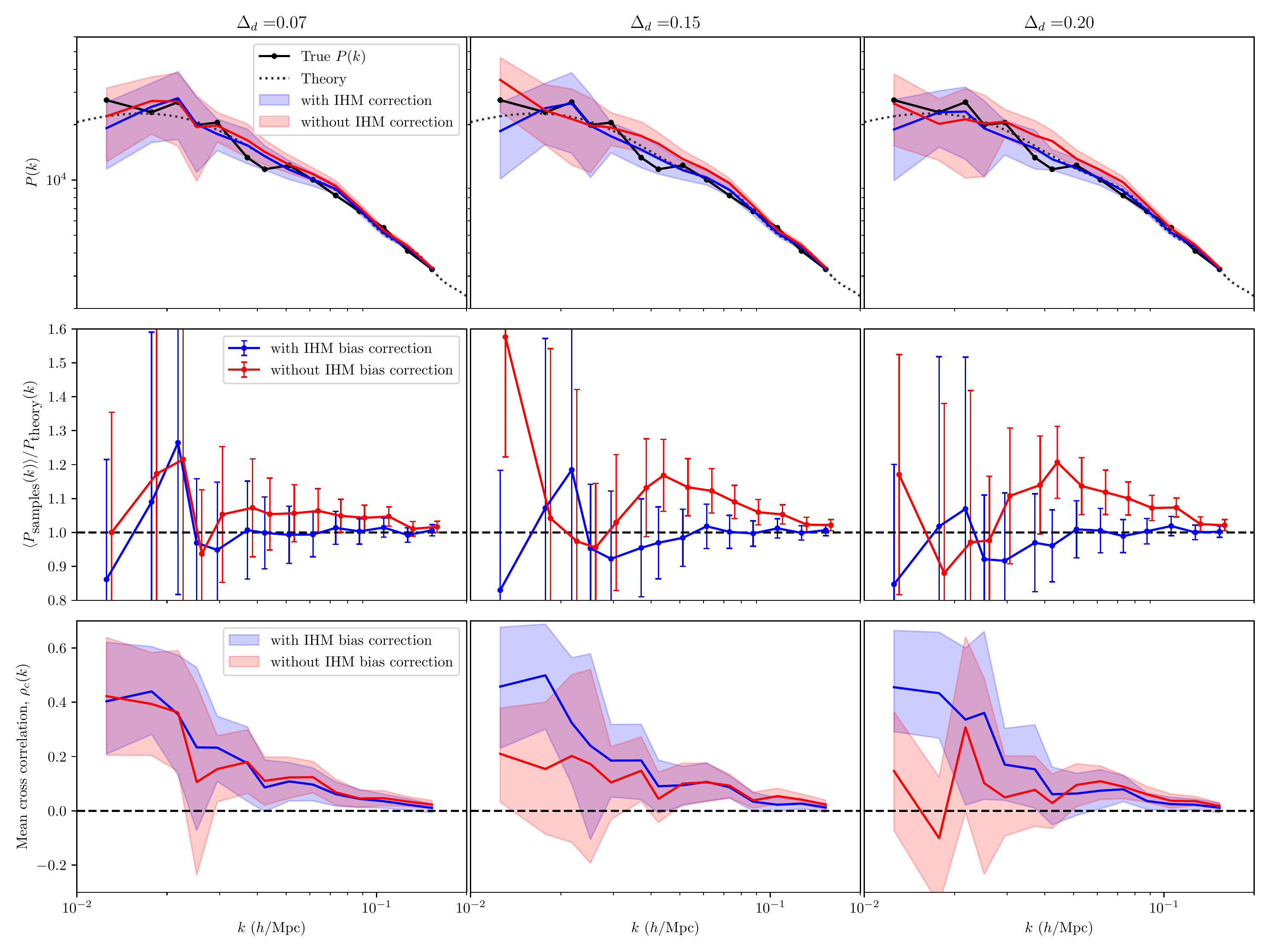}
    \caption{The power spectrum ({\it top}), ratio of the power spectrum to the theoretical expectation ({\it middle}) and the cross-correlation of the reconstructed density field with the true density field ({\it bottom}) for the three mock surveys with distance uncertainties, $\Delta_d=0.07$ ({\it left}), $\Delta_d=0.15$ ({\it centre}) and $\Delta_d=0.20$ ({\it right}). In the top panels, we plot the power spectrum inferred from our reconstruction and compare it with the true power spectrum of the simulation (shown with black line) and the theoretical expectation (shown as a black dotted line). The red lines show the power spectrum of the reconstruction without IHM bias correction and the blue lines are the corresponding curves with IHM bias correction. The shaded region show the $68\%$ confidence interval as calculated from the reconstruction samples. In the middle panels, we plot the ratio of the mean of the power spectrum of our reconstructions to the theoretical expectation. The error bars are calculated assuming a Gaussian covariance for the power spectrum. As we can see the power spectrum of the reconstruction without IHM bias correction is biased high. The magnitude of the bias is higher for the reconstruction with larger distance uncertainty. Finally, in the bottom panel, we plot the cross-correlation of the reconstructed density field with the true density field of the simulation. The cross-correlation is the largest at large scales (small $k$) and it decreases to $0$ at small scales (high $k$). However, we see that if the IHM bias is not corrected, the large scale cross-correlation goes down, suggesting that these density modes are not reconstructed well.}
    \label{fig:summary_stats}
\end{figure*}


\begin{figure*}
    \centering
    \includegraphics[width=\linewidth]{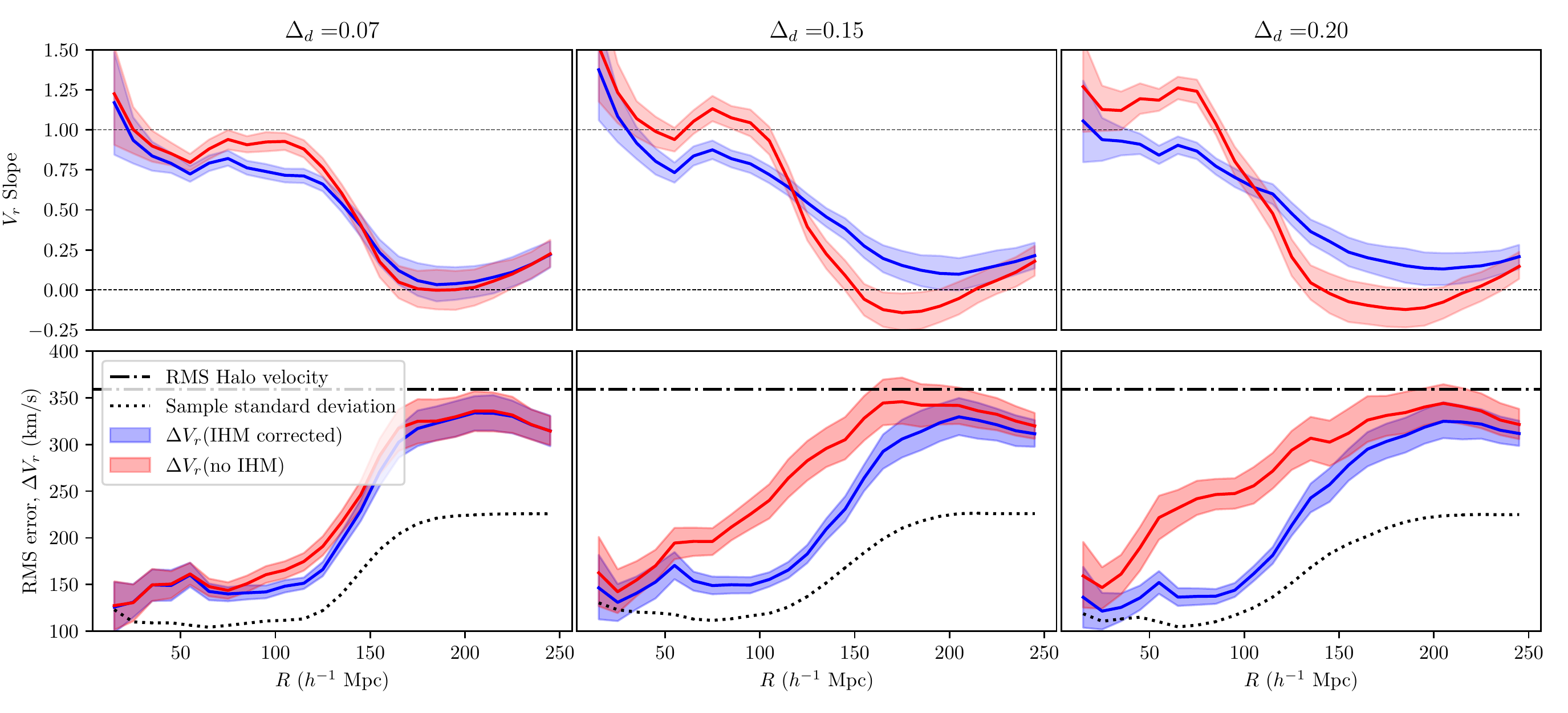}
    \caption{({\it Top}) Slope of regression of the reconstructed $V_r$ on the true $V_r$ values from the simulation, binned in radial shells and ({\it bottom}) the RMS error between the reconstructed velocity and the true velocity. We compare the velocity reconstruction on grid points of the two reconstruction within a given radial shell. The three horizontal panels show the results for mock surveys with three distance uncertainties: $\Delta_d = 0.07$ ({\it left}), $\Delta_d = 0.15$ ({\it centre}) and $\Delta_d = 0.20$ ({\it right}). In the top panel, we plot the slopes of regression in radial shells. The red lines show the results without the IHM bias correction and the blue lines are the results after correcting for the IHM bias. We see that the the slope of regression, $m \sim 1$ deep inside the data region and then drops to $m \sim 0$ at large radius. Without the IHM bias correction, the slope can be however be higher than $1$ since IHM bias introduces spurious velocities. In the bottom panel, we plot the reconstruction error in radial bins. The blue and the red lines show the root-mean-square of the difference of reconstructed and true values of the radial velocity, with and without the IHM bias correction as a function of radius.
    The dotted line shows the standard deviation in our reconstruction samples. The dash-dotted line show the velocity scatter of the halos in the VELMASS simulation. At large radius, the velocity error in our reconstruction approaches the velocity scatter of the halos.
    }
    \label{fig:Vr_pt_estimate}
\end{figure*}

Having introduced our method in the previous section, we now present the results of running our code on a mock peculiar velocity survey. We validate our code by running the code on mock surveys produced from $N$-body simulations and comparing the density and velocity reconstructions with the ground truth.  

\begin{table}
  \centering
  \caption{Depth of the three mock peculiar velocity surveys used in this work. $\Delta_d$ is the fractional uncertainty of the peculiar velocity tracers. $r_{95}$ is the radius of $95\%$ completeness for the mock survey and $r_{\text{peak}}$ is the radius of peak number density of the peculiar velocity tracers.}
  \begin{tabular}{l c c c}
  \hline
      & $\Delta_d$ & $r_{95}~(h^{-1}$ Mpc) & $r_{\text{peak}}~(h^{-1}$ Mpc) \\
  \hline
  Mock1 & $0.07$ & $135$ & $120$ \\
  Mock2 & $0.15$ & $143$ & $117$ \\
  Mock3 & $0.20$ & $176$ & $113$\\
  \hline
  \end{tabular}
  \label{tbl:pv_mocks}
\end{table}
We create our mock peculiar velocity surveys from the VELMASS $N$-body simulation. We select dark matter haloes from the simulation with masses, $M > 3\times 10^{12}~\mathrm{M}_{\odot}$ to avoid using haloes with small number of particles. We then assign an absolute magnitude, $\mM$, to each halo with an scatter, $\sigma_{\text{int}}$. We can estimate the distance to these haloes from their apparent magnitude. Finally, we select the halos brighter than an apparent magnitude, $m_{\text{cut}}$, thus resulting in a magnitude-limited peculiar velocity survey. From the final magnitude-limited sample, we randomly select $7000$ halos. In this paper, we report the results with three different intrinsic scatters, corresponding to distance uncertainties of $7\%$, $15\%$ and $20\%$. As we showed in section \ref{ssec:ihm_bias}, the IHM bias increases for a larger distance uncertainty. Therefore, we anticipate the impact of the IHM bias in our reconstruction to be the larger for the larger distance uncertainty. In Table \ref{tbl:pv_mocks}, we show the depth of the three mock surveys we used in our reconstruction showing the radius of $95\%$ completeness of the mock surveys and the peak of tracer number density.
As mentioned previously, our method requires an estimate of the galaxy density field to correct for the inhomogeneous Malmquist bias. In our runs with the simulations, we use the halo density field of the simulation (smoothed with a Gaussian filter of of scale $4~h^{-1}$ Mpc) as the estimate of the galaxy density field.

To compare the impact of the IHM bias correction, we also ran our method without any IHM correction by setting $\hat{\delta}_g = 0$. As we will see, neglecting the IHM bias correction impacts our reconstruction significantly. A visual comparison of the impact of the IHM correction in the velocity reconstruction is shown in Figure~\ref{fig:visual_comparison}, where we plot the reconstructed velocity field and the associated uncertainty with our algorithm in the $Z=0$ plane. By comparing the reconstructed mean field to the true velocity field in the figure, we can see that our algorithm reconstructs the coherent large-scale velocities well. However, we see that without the IHM bias correction, the reconstructed velocity field has a higher amplitude compared to the true velocity field. This is due to the fact that the IHM bias leads to a spurious velocity in the reconstruction as discussed in section \ref{ssec:ihm_bias}. As expected from equation \eqref{eqn:ihm_bias}, this effect is most pronounced for a distance uncertainty of $\Delta_d=0.20$. Finally, note that outside the data region, the mean velocity field is suppressed to a smaller value of the velocity. 
The impact of the IHM bias is most readily seen in the power spectrum of the reconstructed density field. We show the power spectrum of our reconstruction (with and without the IHM bias correction) in the top panel of  Figure~\ref{fig:summary_stats}. We can see that the reconstructed power spectrum without the IHM bias correction is biased high. This can be seen more clearly in the middle panel of Figure \ref{fig:summary_stats} where we plot the ratio of the mean of the inferred power spectrum to the theoretical expectation. As we can see from the figure, the power spectrum of the reconstruction with IHM bias correction is consistent with the true power spectrum of the simulation. On the other hand, if we do not account for the IHM bias, we see excess power in the reconstructions. This is due to the fact that the IHM bias pushes the velocity to a larger value, which in turn increases the variance in the density field. In order to quantify the bias in the power spectrum, we calculate the quantity 
\begin{equation}\label{eqn:pk_bias}
    \Delta_P = \sum_{i=1}^{N_{k-\text{bins}}}\frac{\langle P_{\textrm{sample}}(k_i) \rangle - P_{\text{theory}}(k_i)}{\sigma[P(k_i)]},
\end{equation}
where $\sigma[P(k_i)]$ is the standard deviation in the power spectrum assuming a Gaussian covariance. For an unbiased reconstruction, we expect this quantity to have a value of zero. We compare the bias in the reconstruction by computing the bias in the power spectrum measured in $15$ bins in $k$. For the reconstruction with the IHM bias correction, we obtain values of $\Delta_P = +0.46, -0.25, +0.55$ for the mock surveys with $\Delta_d = 0.07, 0.15, 0.20$ respectively. The same quantities when measured for the reconstruction without the IHM bias correction are $\Delta_P = +8.78$, $+15.24$ and $+16.99$ respectively showing that the power spectrum inferred without IHM bias correction is biased high with high significance. 

While the power spectrum gives a useful consistency check, the power spectrum does not provide an assessment about the phases of the Fourier modes. In order to test that the phases of the density modes are reconstructed correctly, we compute the cross-correlation rate of the reconstructed density field with the true underlying density field, which is defined as follows,
\begin{equation}
    \rho_c(k) = \frac{\langle \delta_{\text{rec}}(\mvec{k}) \delta^{*}_{\text{sim}}(\mvec{k})\rangle_{\mvec{k} \in k~\text{bin}}}{V_s\sqrt{P_{\text{rec}}(k) P_{\text{sim}}(k)}},
\end{equation}
where $P_{\text{rec}}$ is the power spectrum of the reconstruction and $P_{\text{sim}}$ is the true power spectrum of the simulation. $V_s$ is the volume of the survey. We show the cross-correlation of our reconstructed density samples with the true underlying density field of the simulation in the bottom panels of  Figure~\ref{fig:summary_stats}. As can be seen, the reconstructed large scale density modes shows a cross-correlation of $\sim 0.5$ with the true density field, showing that the large scale density modes are reconstructed well with our method. The cross-correlation approaches zero for the small-scale ($k \gtrsim 0.04~h$~Mpc$^{-1}$) modes. We can see the impact of the IHM bias on the cross-correlation from this plot. For the mock survey with larger distance uncertainty, for the reconstruction without IHM bias correction, cross-correlations of the large-scale density modes are reduced compared to the reconstruction with IHM bias correction. This effect is most readily seen in the reconstruction with $\Delta_d = 0.2$

We however note that, while the cross-correlation in Fourier space is an interpretable metric, it may not be the best metric to compare the two fields. This is because we only have the peculiar velocity data in a small volume of our simulation box. Therefore, we also compare the radial velocity in our reconstruction to the true radial velocity in shells of radial distances. In each shell, we compute the slope, $m$ and the scatter, $\sigma_v$ between the velocity estimates by regressing the reconstructed velocity on the true velocity field. We use this slope as a proxy for the cross-correlation of the two fields in real space. We show the result from this comparison in Figure \ref{fig:Vr_pt_estimate}. As it can be seen from the figure, IHM bias corrected reconstruction has a slope of $m \sim 1$ in the data region ($r \lesssim 120~h^{-1}$ Mpc) and then decreases to a value of $m \sim 0$ at large $R$. On the other hand, the IHM bias leads the reconstructed velocity to be biased high, which is reflected by the fact that the slope, $m$ can be higher than $1$ for the reconstructions with large IHM bias. Furthermore, we compare the reconstruction error for the different runs of our method. The reconstruction error is defined as the Root mean squared (RMS) error between the reconstruction and the true velocity, $(\Delta V_r)^2 = |V^{\text{samples}}_r - V^{\text{true}}_r|^2$. We see that the IHM bias corrected reconstruction leads to a smaller error in the reconstructed velocity compared to the reconstruction without IHM bias correction. At large radius, the velocity scatter between the reconstructed velocity and the true velocity approaches the scatter of the halos in the VELMASS simulation. We discuss in more detail the different sources contributing to this reconstruction error in Appendix \ref{app:error_origin}.

The above tests show that we get an unbiased reconstruction of the velocity field with our forward-modelled reconstruction method after correcting for the IHM bias. On the other hand, not correcting for the inhomogeneous Malmquist bias can substantially bias the velocity field reconstruction. We now move to use our reconstruction method on real peculiar velocity surveys and reconstruct the velocity field of the local Universe.

%% file: sections/real_data.tex
\begin{figure}
    \centering
    \includegraphics[width=\linewidth]{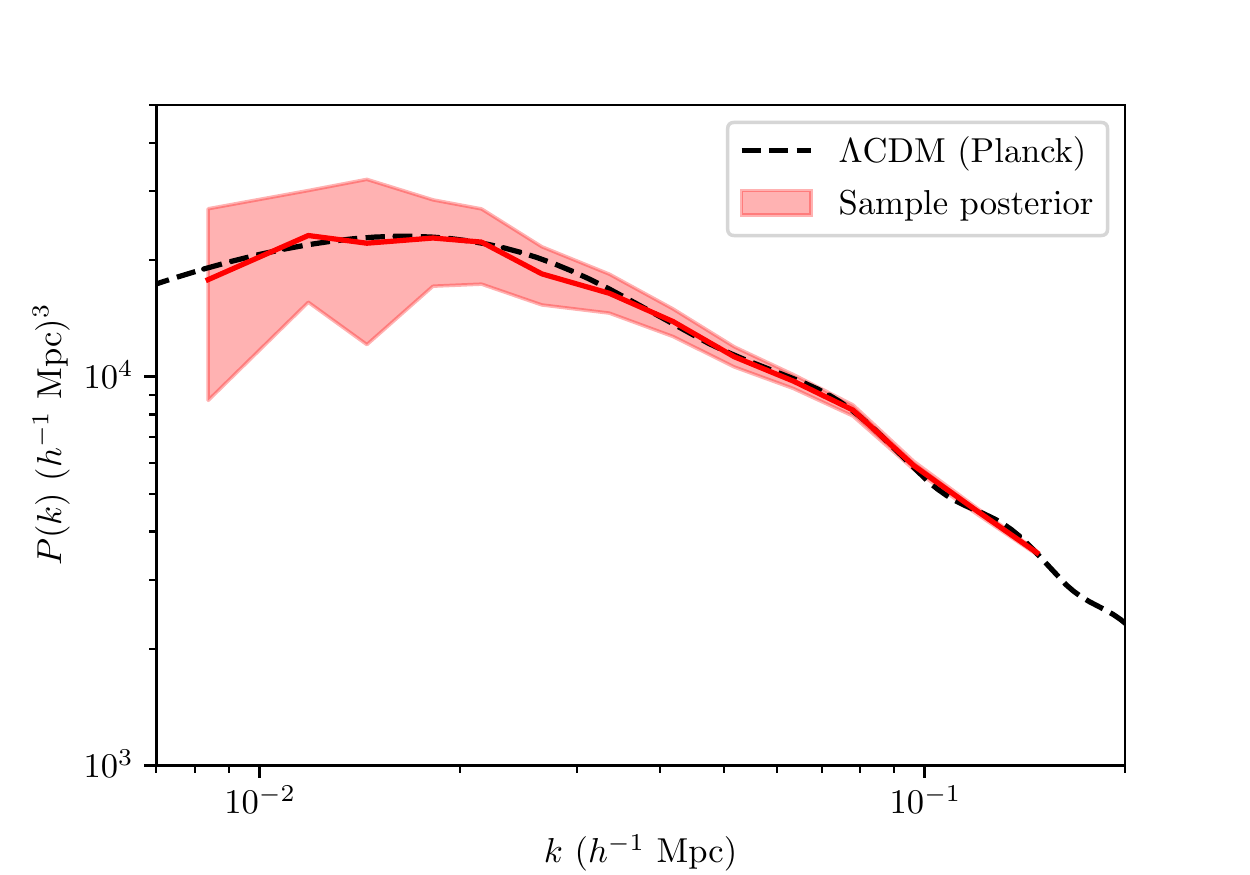}
    \caption{Power spectrum of the samples of reconstruction of the local Universe. The reconstruction was performed using the 2MTF and the SFI++ Tully-Fisher catalogues and the A2 supernovae compilation. The density reconstruction from \citet{Carrick_et_al} was used for the IHM bias correction.}
    \label{fig:Pk_real_data}
\end{figure}

\begin{figure*}
    \centering
    \includegraphics[width=\linewidth]{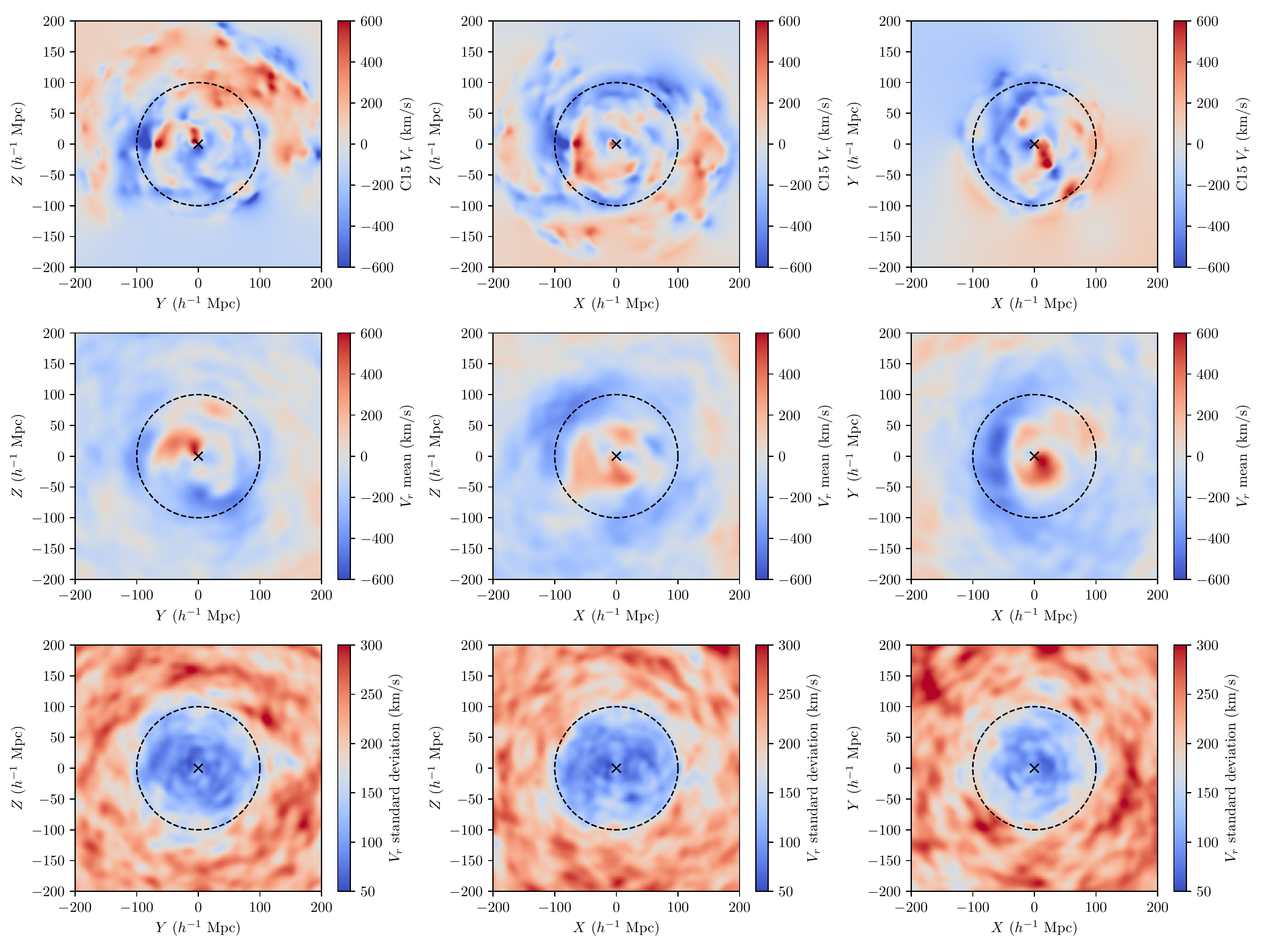}
    \caption{A comparison of our reconstruction with the reconstruction of \citet{Carrick_et_al} in the galactic $X$ ({\it left}), $Y$ ({\it centre}) and $Z$ ({\it right}) planes. The top panel show the reconstruction of \citet{Carrick_et_al}, the middle and the bottom panels show the mean and the standard deviation of the velocity reconstruction from our method. We can see similar large scale features in the velocity field in both our reconstruction and the C15 reconstruction. We show the radial distance of $100~h^{-1}$ Mpc with a black dotted circle. This is the distance cut we use for our peculiar velocity tracer sample. Note that the uncertainty in the velocity field reconstruction increases drastically beyond this boundary.}
    \label{fig:2mpp_comparison}
\end{figure*}

\section{Velocity reconstruction in the local Universe}\label{sec:vel_rec}

\begin{figure*}
    \centering
    \includegraphics[width=\linewidth]{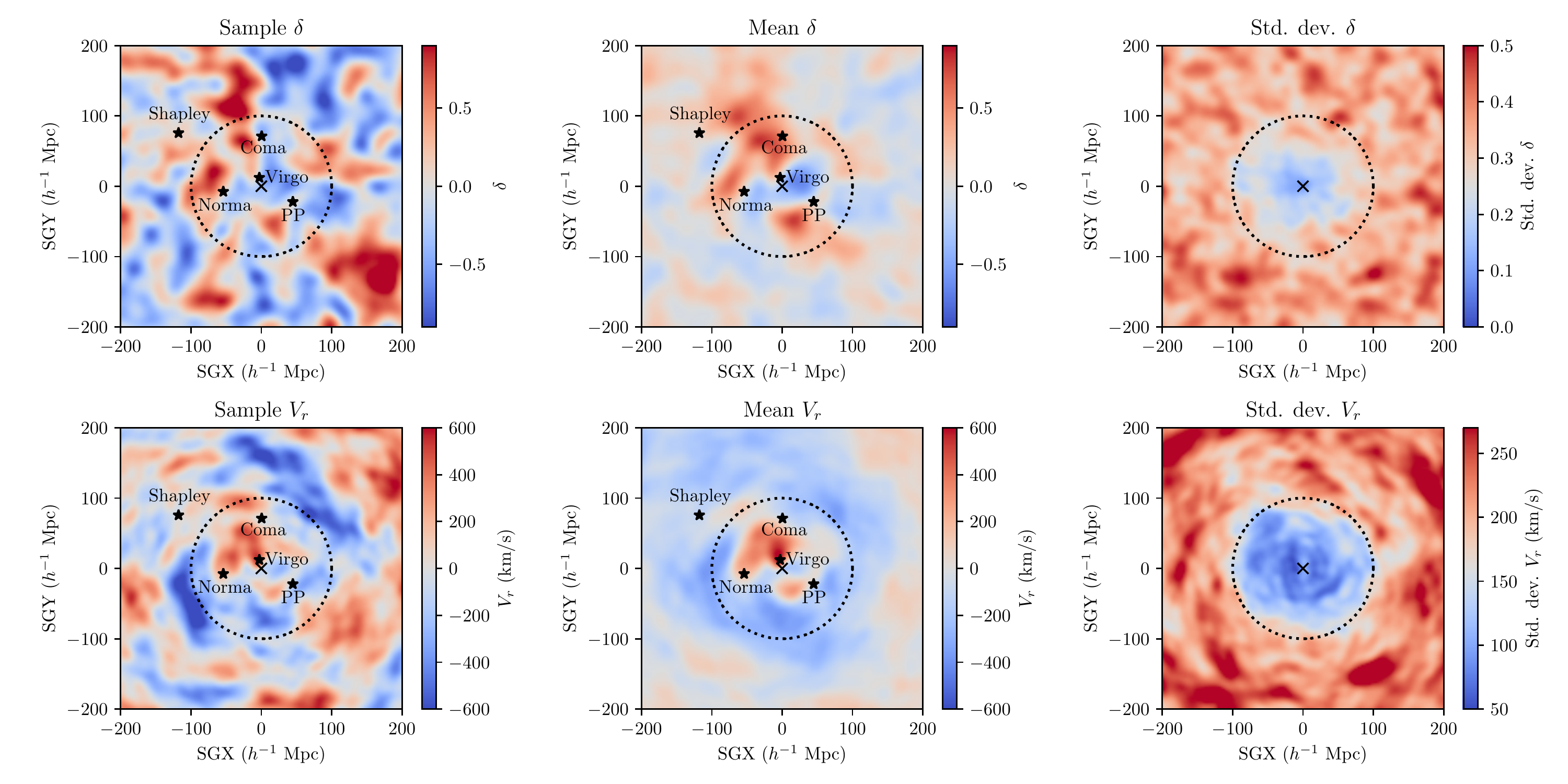}
    \caption{The density and the velocity field reconstruction in the supergalactic plane. The top plot shows the density field smoothed using a Gaussian filter of smoothing scale $10~h^{-1}$ Mpc and the bottom panel shows the velocity field reconstruction. On the left panel we show the density and the velocity field of a randomly selected sample in our reconstruction. The middle panels show the mean density and velocity field. The rightmost panels show the standard deviation in our density and velocity field reconstruction samples. We show the location of several prominent clusters in the supergalactic plane, namely, Virgo, Norma, Perseus-Pisces (PP), Coma and Shapley, with a black star. The black dotted line shows the radius at $100~h^{-1}$ Mpc, which is the selection limit imposed on the sample.}
    \label{fig:sg_plot}
\end{figure*}
\begin{figure*}
    \centering
    \includegraphics[width=0.8\linewidth]{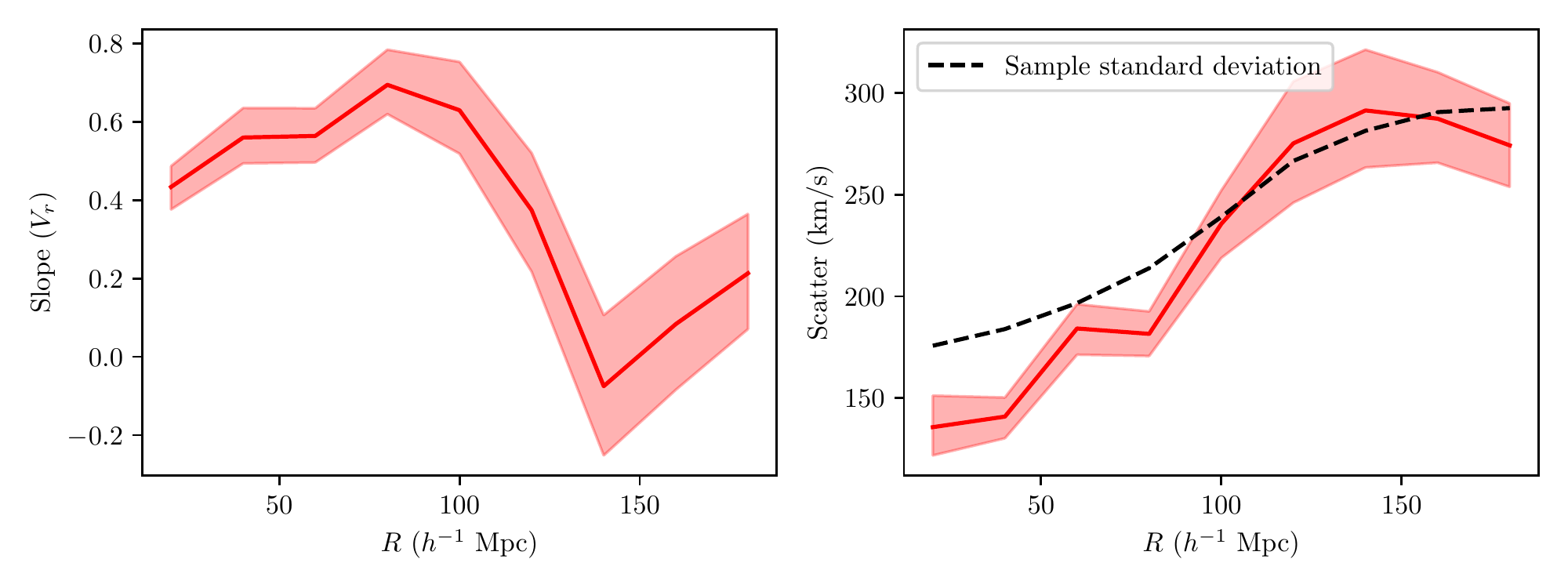}
    \caption{Comparison of ({\it left}) the slope and ({\it right}) the RMS difference between the velocity estimates of our reconstruction and the C15 reconstruction for voxels binned in radial shells. The slope can be interpreted as the cross-correlation between the two velocity fields. We see a large degree of cross-correlation in the region with the peculiar velocity data ($R \lesssim 100~h^{-1}$ Mpc), while it rapidly goes to zero beyond this region. The right panel shows the root-mean-squared error between the two velocity reconstructions. At very small radius, $R \lesssim 50 h^{-1}$ Mpc, the scatter approaches the value, $\sigma_{\text{NL}} \approx 150$ km/s. The black dashed line show the standard deviation in our reconstruction sample with $\sigma_{\text{NL}}$ added in quadrature.}
    \label{fig:Vr_pt_comparison}
\end{figure*}

\begin{figure*}
    \centering
    \includegraphics[width=\linewidth]{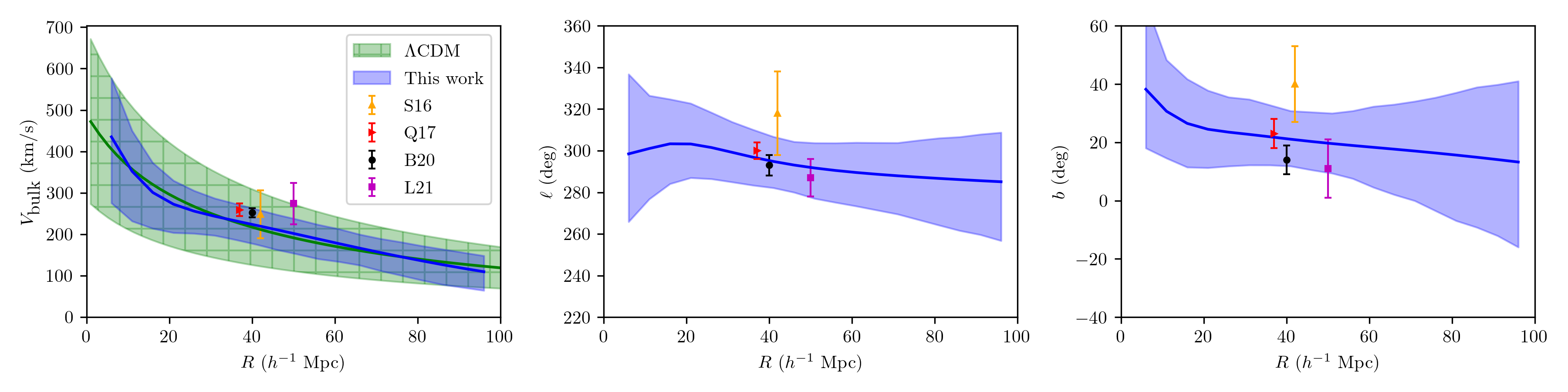}
    \caption{Comparison of the bulk flow measured in our reconstruction and compared to the $\Lambda$CDM expectation as well as the other results in the literature. The left panel shows the magnitude of the bulk flow, while the middle and the right panel shows the direction of the bulk flow in terms of the galactic longitude and latitude respectively. As we can see, the magnitude of the bulk flow in our reconstruction (shown in blue) is consistent with both the $\Lambda$CDM expectation (shown in green hatch) as well as other results in the literature. The direction of the bulk flow as inferred from our reconstruction is also consistent with other results in the literature. The results in the literature compared in this results are: \citet{Scrimgeour2016} (S16), \citet{Qin2019} (Q17) \citet{Boruah2020} (B20), \citet{Lilow2021} (L21)}
    \label{fig:bulk_flow}
\end{figure*}

In this section, we apply our method on real peculiar velocity data to reconstruct the velocity field of the local Universe. 

We use the SFI++ and 2MTF Tully-Fisher catalogues and the A2 supernovae data set in this work. We use the galaxy density reconstruction of \citet{Carrick_et_al} to correct for the inhomogeneous Malmquist bias. Since this reconstruction is limited to $125~h^{-1}$ Mpc in certain sky directions, we only consider peculiar velocity tracers within estimated distances, $d < 100 h^{-1}$ Mpc in this work. Note that, this does not lead to selection effects in the forward methods of peculiar velocity analysis \citep{StraussWillick95}. We marginalize over a scaling factor to account for uncertainties in the zero point calibration. We use a different scaling factor for each peculiar velocity survey. By sampling density fields from the posterior \eqref{eqn:posterior}, we get samples of reconstruction that are consistent with the observed data. One useful side product of sampling from the posterior is that we get an estimate of the correlated uncertainties in the reconstructed velocity field. For the reconstruction of the local Universe, we use a cubic sampling box of side length, $L_{\text{box}} = 500~h^{-1}$ Mpc with $128^3$ grids. In Figure \ref{fig:Pk_real_data}, we plot the power spectrum of the samples of our reconstruction. Since we assume a Gaussian prior on the density field, we expect the reconstruction to have a cosmological power spectrum. However, as we saw in section \ref{sec:sim_results}, IHM bias may artificially enhance the inferred power spectrum. Therefore, consistency between the theoretical power spectrum and the inferred power spectrum shows that the IHM bias is corrected in our reconstruction. As can be seen from the figure, this is indeed true and the power spectrum computed from the samples of our posteriors agree very well with the theoretical prediction from the $\Lambda$CDM model.

\subsection{Comparison with 2M++ reconstruction}
Unlike the simulations, for the real data, we do not have access to the true velocity field. Nonetheless, we can compare our reconstruction to velocity fields reconstructed using other methods. Therefore, we compare our reconstructed velocities to the reconstruction of \citet{Carrick_et_al} who used an iterative reconstruction procedure, where the galaxy density is reconstructed from the galaxies in the 2M++ galaxy compilation \citep{Lavaux2011}. The velocity is predicted from the galaxy density field using linear perturbation theory. However, since galaxies are a biased tracer of the underlying density field, the velocity estimate from the galaxy field needs to be scaled by a factor of $\beta = f / b$. The value of $\beta$ is then fitted by comparing the predicted velocities to the the peculiar velocity data from a peculiar velocity survey. In this work, we use the best fit $\beta$ and $\mvec{V}_{\text{ext}}$ value fitted in \citet{Carrick_et_al}. We show a visual comparison of our velocity reconstruction with the reconstruction of \citet{Carrick_et_al} (hereafter C15) in the galactic $X, Y, Z$ planes in Figure \ref{fig:2mpp_comparison}. The top panels of Figure \ref{fig:2mpp_comparison} shows the radial velocity field in the reconstruction of C15. The middle and the bottom panels show the mean and the standard deviation of the velocity field samples reconstructed using our algorithm. By comparing our velocity reconstruction  to the 2M++ reconstruction, we see the similarity between some of the same large scale features in both reconstruction. Note that in the region without data, the mean field approaches very small values. This is similar to the Weiner filtered fields, where in the region without data, the field is strongly suppressed such that in the limit of no data, the velocity field approaches zero. In the bottom panel, we plot the uncertainty in the velocity estimates in these planes. Since our method produces samples of reconstruction from the posterior with the Gaussian prior, we get the correlated uncertainties in velocity estimates of our reconstruction. From the figure, we see that the uncertainties in the velocity estimates is substantially lower in the inner region which has the peculiar velocity data than the outer region which does not have peculiar velocity data. We also plot the velocity field and the density field in the supergalactic plane comparing also the position of some prominent galaxy clusters in Figure \ref{fig:sg_plot}. From the mean field, we see that while the over-densities around the Perseus-Pisces, Virgo and Coma clusters are well-reconstructed, we do not see prominent overdensities around the Norma and the Shapley clusters. Since the Shapley cluster lies outside the $100~h^{-1}$ Mpc, the threshold of our reconstruction and the Norma cluster lies along the galactic plane, there is lack of peculiar velocity data around these clusters. The lack of peculiar velocity data around the cluster leads to poor reconstruction quality around these regions.

Finally, we compare our reconstruction to the C15 reconstruction by comparing the velocity fields in radial shells. For this comparison, we compared the radial velocity from C15 and our reconstruction in radial shells of width $20~h^{-1}$ Mpc. In each shell, we compare the radial velocities from our reconstruction samples and the reconstruction from C15. We fit the slope and the standard deviation between the two velocity estimates by regressing the velocity estimates of our reconstruction on the C15 velocity estimates. The fitted slope can be thought of as a proxy for the cross-correlation between the two velocity estimates. The slope and the scatter between the two velocity fields are shown in Figure \ref{fig:Vr_pt_comparison}. As we can see from the figure, the two velocity fields have a large degree of correlation in the nearby Universe ($R \lesssim 120~h^{-1}$ Mpc). Since the peculiar velocity data is restricted to $d < 100 h^{-1}$ Mpc, we do not expect the reconstructed velocity field to be a good estimate beyond this boundary. This is reflected in the fact that the slope rapidly approaches zero beyond the peculiar velocity data boundary. We also see that the scatter between the two increases with increasing radius. At very low radius ($R \lesssim 50~h^{-1}$ Mpc), the scatter between the two velocity fields is less than or comparable to the uncertainty due to the non-linearities, $\sigma_{\text{NL}} \approx 150$ km/s. At large radius, the scatter approaches the sample standard deviation ($\sim 250$ km/s).

\subsection{Bulk flow}

\begin{table*}
  \centering
  \caption{Comparison of the bulk flow results with other studies in the literature. The results are also shown in Figure \ref{fig:bulk_flow}. We see that our bulk flow result is in excellent agreement with the other results in literature.}
  \begin{tabular}{l|l|l|c|c|c|c}
  \hline
    Reference & Abbreviation & Effective radius  & $|\mvec{V}_{\textrm{bulk}}|$ (km/s) & $l$ (deg) & $b$ (deg) \\
    \hline
   \citet{Scrimgeour2016} & S16 & $50\ h^{-1}$ Mpc & $248 \pm 58$ & $318 \pm 20$ & $40 \pm 13$\\
   \citet{Qin2019} & Q17 & $37\ h^{-1}$ Mpc & $259 \pm 15$ & $300 \pm 4$ & $23 \pm 3$\\
   \citet{Boruah2020} & B20 & $40\ h^{-1}$ Mpc & $252\pm11$ & $293 \pm 5$ & $14 \pm 5$ \\
   \citet{Lilow2021} & L21 & $50\ h^{-1}$ Mpc & $274\pm50$ & $287 \pm 9$ & $11 \pm 10$ \\
   \textbf{This work} & --- & $\boldsymbol{40\ h^{-1}}$ \textbf{Mpc} & $\boldsymbol{220\pm21}$ & $\boldsymbol{295 \pm 6}$ & $\boldsymbol{21 \pm 5}$ \\
    \hline 
  \end{tabular}
  \label{tbl:bulk_flow}
\end{table*}

Due to the sensitivity to large scale density modes, the velocity field is correlated on very large scales. 
The large scale velocity field is often expanded in terms of its kinematic moments, such as the dipole and the quadrupole of the velocity field \citep{Jaffe1995},
\begin{equation}
    \mvec{V}(d) = \mvec{U} + d \mmat{Q}\cdot \hat{\mvec{r}} + ...,
\end{equation}
where, $\mvec{U}$ is the dipolar bulk flow and $\mmat{Q}$ is the trace-free, symmetric quadrupolar shear moment and $d$ is the radial distance. The bulk flow is sensitive to the large scale power of the density field and therefore can also be used to constrain cosmological parameters due to its sensitivity to the power spectrum shape parameter \citep{Feldman08}.

We can measure the bulk flow and the shear moments directly from our reconstruction. In order to measure the bulk flow in our reconstructions at a given scale, we smooth the velocity field with a window function of characteristic scale, $R_{\text{bulk}}$. We use a Gaussian filter in order to measure the bulk flow in our reconstruction. The bulk flow at a scale of $R_{\tbulk} = 40 h^{-1}$ Mpc for our reconstruction is $V_{\tbulk} = 220^{+20}_{-21}$ km/s in the direction $l=295^{\circ} \pm 6^{\circ}$, $b=21^{\circ} \pm 5^{\circ}$ degree. The bulk flow as measured in our reconstruction is compared to a number of different results in the literature Table \ref{tbl:bulk_flow}. We also show the total magnitude and the direction of the bulk flow as measured from our reconstruction samples in Figure \ref{fig:bulk_flow}. As we can see from the figure, both the magnitude and the direction of the bulk flow is consistent with other results in the literature. Of the compared results, \citet{Scrimgeour2016} and \citet{Qin2019} use an estimator to estimate the bulk flow directly from the peculiar velocity data. On the other hand, \citet{Boruah2020} and \citet{Lilow2021} explicitly fit an external dipole in their flow model to account for velocity contribution from outside the survey volume. Note that the approach presented here is different from these approaches. We do not explicitly model an additional dipole. The density modes we fit for in our reconstruction not only accounts for the bulk flow, but also other higher kinematic moments of the large scale velocity field.


\subsection{Velocity field comparison}

\begin{table}
  \centering
  \caption{Comparison of the logarithm of Bayes factor for various redshift selection. A positive log of Bayes factor implies that our model is favored over the compared model, while a negative log for the Bayes factor implies the compared model is preferred over our model. From the table, we can see that our model performs better than the adaptive kernel smoothed velocity field. On the other hand C15 velocity field performs better than our forward modelled velocity field.}
  \begin{tabular}{l l c c c}
  \hline
     Test set & Selection & $\ln\bigg(\frac{\mP_{\text{\text{fwd-PV}}}}{\mP^{\text{TF}}_{\text{adaptive}}}\bigg)$ &     
     $\ln\bigg(\frac{\mP_{\text{fwd-PV}}}{\mP_{\text{2M++}}}\bigg)$ &$N_{\text{tracers}}$\\
    \hline
    A2 &  $cz < 0.01$ & $7.44$ & $-2.12$ & $49$ \\
    & $z < 0.015$ & $10.88$ & $-5.92$ & $92$ \\
    & $z < 0.02$ & $21.73$ & $-20.69$ & $168$ \\
    & $z < 0.03$ & $41.16$ & $-27.65$ & $310$ \\
    \hline
  \end{tabular}
  \label{tbl:model_comparison}
\end{table}

We introduced a Bayesian model comparison framework in \citet{Boruah2021} to compare the performance of different velocity reconstruction models. In this model comparison framework, we look at the Bayes factor between two models, $\mM_1$ and $\mM_2$, 
\begin{equation}
    \text{Bayes factor} = \frac{\mP(\mD|\mM_1)}{\mP(\mD|\mM_2)}.
\end{equation}
If the Bayes factor is greater than 1, the model $\mM_1$ is preferred over the model $\mM_2$ and vice-versa. We use this model comparison framework to assess the quality of the velocity reconstruction method introduced in this paper. We compare our velocity model to two different velocity reconstruction of the local Universe - {\it i)} The reconstruction of C15, {\it ii)} an adaptive kernel-smoothed velocity reconstruction, where the peculiar velocity data is directly smoothed \citep{Springob2014, Springob2016}. We use an adaptive-kernel smoothed velocity field obtained from a combined Tully-Fisher catalog consisting of the SFI++ and the 2MTF velocity field. The Bayesian model comparison method relies on our capacity to predict new data points from the model. This is called a posterior predictive test, and thus requires what is sometimes called a ``test set'' of peculiar velocities. Therefore for this comparison, we perform a reconstruction using only the Tully-Fisher catalogs for the reconstruction and use the A2 sample as a test set for the posterior predictive analysis. We note that this is the same peculiar velocity data set used for the adaptive kernel-smoothed reconstruction. 

The results of the model comparison is shown in Table \ref{tbl:model_comparison}. Using our model comparison comparison, we find that the forward-modelled velocity reconstruction performs better than the kernel smoothed velocity field, even though both reconstructions use the same data sets. Our velocity field however does not perform as well at predicting new velocity data than the reconstruction of \cite{Carrick_et_al}, which uses the 2M++ galaxies to map the density field, and linear perturbation theory to predict peculiar velocities. This is not a failure of our method, but rather a limitation of the peculiar velocity data used for the reconstruction. Since the peculiar velocity data used in this work are very noisy and sparse, the quality of reconstruction is not comparable to the 2M++ reconstruction. In the future, with denser peculiar velocity samples, we expect the quality of reconstruction to be significantly better. There is also the possibility that the forward linear modeling approximation may fail which would require more complex models based, e.g., on \borg. 

%% file: sections/discussion.tex
\section{Discussion}\label{sec:discussion}

Our work provides a way towards including peculiar velocity data into the initial condition reconstruction framework, {\sc borg}. A density reconstruction by combining both galaxy and peculiar velocity data in {\sc borg} potentially has multiple advantages. First, using \borg, we can access the fully non-linear velocity field, which is accurate to much smaller scales as compared to the velocity field from linear perturbation theory. For example, \borg particle mesh runs has been shown to perform better than linear velocity fields in $N$-body simulations \citep{Mukherjee2021}. Furthermore, the vorticity of the velocity field, which is a uniquely non-linear phenomenon was already reconstructed in \citet{Jasche2019}. Second, adding the velocity data to the forward-modelled reconstruction can potentially make the reconstruction robust to the details of the galaxy bias model. As was shown in \citet{Nguyen2021}, the amplitude of the inferred power spectrum in \borg reconstruction with galaxy/halo catalogues alone may be substantially biased. This arises because the higher order bias terms are not sufficiently informative to break the degeneracy between linear galaxy bias and the amplitude of matter clustering, $\sigma_8$. Since peculiar velocity data is sensitive to the total matter density, the addition of peculiar velocity data to galaxy data would make the \borg~reconstruction robust to the details of the galaxy bias. Adding velocity information to the forward-modelled reconstruction can potentially lead to improved reconstruction of the full phase space structure (density + velocity) of dark matter \citep{Leclercq2017}. Such phase space reconstruction can then potentially guide observation efforts for new discoveries \citep{Kostic2021}.  

In the future, we can also extend our method to simultaneously constrain cosmological parameters from peculiar velocity data. As we highlight in this paper, not accounting for the IHM bias can lead to the amplitude of the peculiar velocities and so the inferred power spectrum of the density/velocity field to be biased high. This is crucial for the inference of the cosmological parameters, since this bias will likely translate to the biased inference of cosmological parameters. Thus, this paper provides the foundation of inferring cosmological parameters with peculiar velocity data in a forward modelled framework. Also note that our method is sensitive to the large-scale flows, which in turn are a sensitive probe of the break in the matter power spectrum \citep[e.g,][]{Feldman08}. Thus, the large-scale power spectrum can constrain the shape parameter, $\Gamma\approx\Omega_m h$. Combining these constraints from large-scale flows with traditional velocity-velocity comparison method can potentially provide additional cosmological information. The use of the non-linear velocity field using \borg~can also potentially break the $f\sigma_8$ degeneracy through the non-linearities. This feature was already noted in the context of weak lensing by using a forward model based on Lagrangian Perturbation theory within the \borg framework for mock weak lensing data \citep{Porqueres2021b}.

Another possible application of our forward-modelled velocity reconstructions is in using the kinetic Sunyaev-Zeldovich (kSZ) data to reconstruct the largest scale velocity modes. kSZ-based velocity reconstruction has been shown to improve the constraints on primordial non-Gaussianities through multi-tracer analyses \citep{Giri2020}. 

%% file: sections/summary.tex
\section{Summary}\label{sec:summary}

In this paper, we introduced a forward-modelled velocity reconstruction method which uses peculiar velocity data to reconstruct the velocity field. Using an external estimate of the line of sight density, we consistently correct for the inhomogeneous Malmquist (IHM) bias. Using mocks created from an $N$-body simulation, we validated that our method leads to unbiased velocity field reconstruction after accounting for the inhomogeneous Malmquist bias.  However, as we show using mock simulations, not accounting for the inhomogeneous Malmquist bias in forward-modelled methods can lead to significantly biased reconstruction. Since the IHM bias may induce spurious flows, without the correction for the IHM bias, the inferred power spectrum is biased high. We then applied our method to the 2MTF, SFI++ Tully-Fisher catalogues and the A2 supernovae compilation, resulting in a novel forward-modelled velocity field reconstruction of the local Universe. Since we sample our velocity field reconstruction samples from the field-level posterior, we get an estimate of the full correlated uncertainties in the peculiar velocity estimates. Furthermore, we found that the bulk flow calculated from our reconstruction is consistent with other results in the literature. Using a Bayesian model comparison framework, we showed that the reconstructed forward modelled velocity fields perform better than the widely used adaptive kernel-smoothed velocity fields constructed from the same data. However, the reconstruction presented here does not perform as well as the velocity field reconstructed from the 2M++ galaxy catalogue. This is likely because of peculiar velocity data sets are sparser and noisier compared to the galaxy catalogues. Our method provides a way to extend forward-modelled initial condition reconstruction algorithms such as \borg~by including peculiar velocity data sets. 

%% file: sections/appendix/velocity_error.tex
\section{Physical origin of the reconstruction error}\label{app:error_origin}

\begin{figure}
    \centering
    \includegraphics[width=\linewidth]{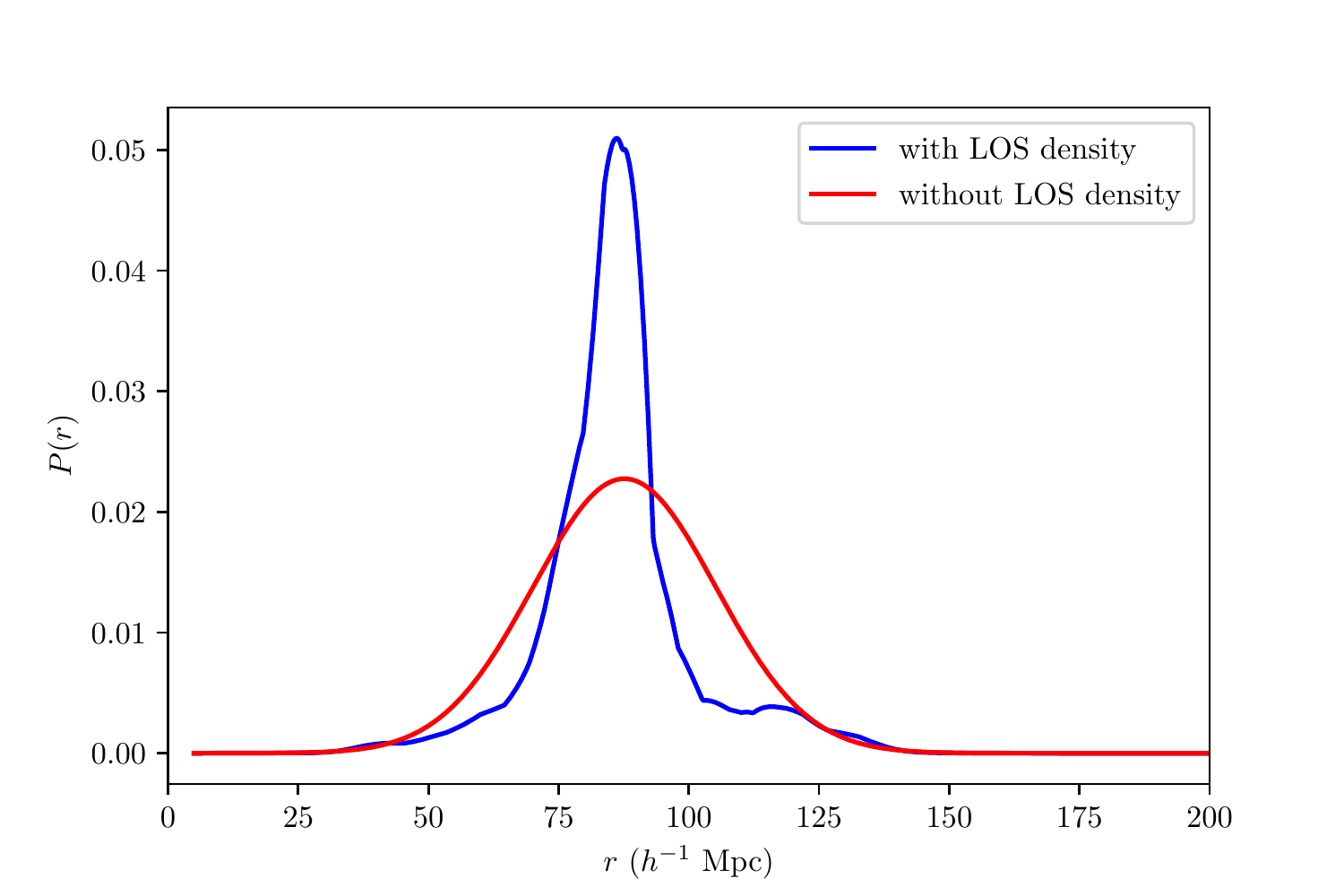}
    \caption{The impact of the line-of-sight density field on the radial distribution of the peculiar velocity tracer. Here we show the radial distribution of a peculiar velocity tracer in our mock survey. The red curve shows the expected radial distribution ignoring the line-of-sight inhomogeneities and the blue curve shows the distribution after accounting for the LOS inhomogeneities. As we can see, radial distribution after accounting for the line-of-sight inhomogeneities is more peaked than the naive distribution expected from the measured distance error.}
    \label{fig:los_density}
\end{figure}

In this appendix, we try to isolate different physical origins of the uncertainties in our velocity reconstruction. As shown in section \ref{sec:sim_results}, without the use of the correct line-of-sight (LOS) distribution for the peculiar velocity tracers, the reconstructed velocity field is biased high. This bias also leads to increased uncertainty in the reconstructed velocity. However, another factor contributes to the increased error in the velocity reconstruction. In equation \eqref{eqn:los_distribution}, there are two contributions to the distribution of peculiar velocity tracers along the LOS - {\it i)} The uncertainty in the distance, {\it ii)} The line of sight density field. Due to the latter contribution, peculiar velocity tracers are clustered around regions of high density. Therefore, when we take the LOS density into account, the radial distribution may be more peaked than naively expected from the measured distance uncertainty. This is shown in Figure \ref{fig:los_density}, where we plot an example from our mocks used in section \ref{sec:sim_results}. In this figure, the radial distribution of the tracer including the LOS density is much sharply peaked compared to the distribution is we ignored the LOS.

\begin{figure}
    \centering
    \includegraphics[width=\linewidth]{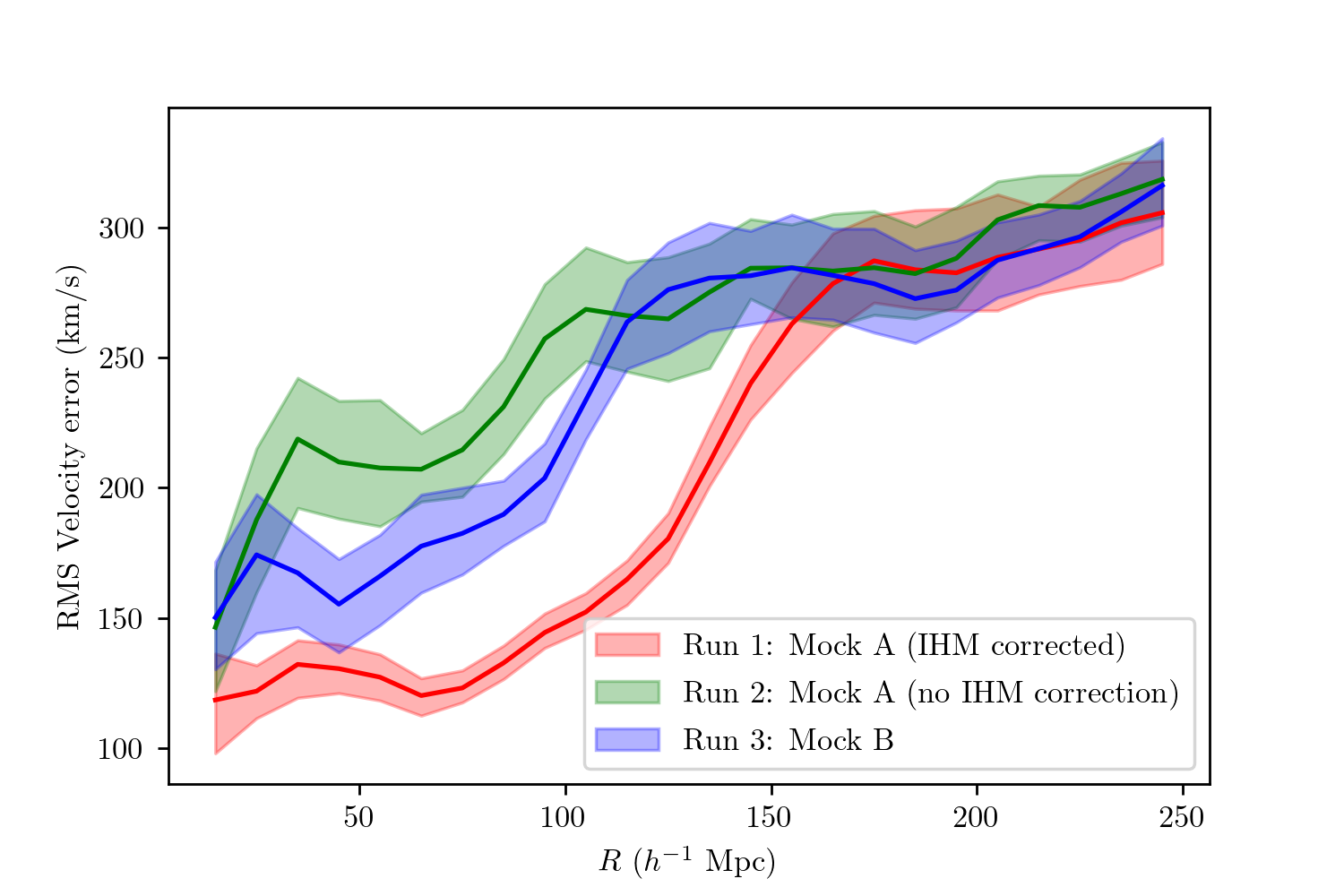}
    \caption{The RMS error in the velocity reconstructions in our three runs. The peculiar velocity tracers in Mock A are sampled by accounting for the LOS inhomogeneities and therefore the reconstruction is susceptible to IHM bias. On the other hand, in Mock B, the peculiar velocity tracers are sampled homogeneously and therefore does not suffer from IHM bias. We notice that the RMS error in Run 1 is much lower than Run 2 and 3. On the other hand, the RMS error in Run 3 is closer to the error in Run 2 (compared to Run1). This suggests that the error in the IHM corrected reconstruction has the additional advantage of reducing the reconstruction error by better localization of the peculiar velocity tracers.}
    \label{fig:velocity_scatter}
\end{figure}

Therefore, when we account for the LOS density, we not only correct for the IHM bias, but we also sharpen the distance estimate and therefore decrease the uncertainty in the reconstruction. In order to investigate the impact of this factor, we use our method on two different mock surveys. In mock survey A, we populate the tracers according to equation \eqref{eqn:los_distribution}, accounting for the LOS density field. In mock B, we populate tracers homogeneously, i.e, we do not account for the LOS inhomogeneity. Therefore, by design, inference with mock B does not suffer from IHM bias. Since the velocity tracers are not clustered around density peaks, we are not able to determine the position of the tracer better than the distance uncertainty. We then make three runs of our code on the two mocks:
\begin{itemize}
    \item Run 1: Mock A, corrected for IHM bias.
    \item Run 2: Mock A, not corrected for IHM bias.
    \item Run 3: Mock B, not corrected for IHM bias (by design, Mock B is unaffected by IHM bias).
\end{itemize}
The results of the three runs are shown in Figure \ref{fig:velocity_scatter}, where we plot the Root Mean Squared (RMS) error in the velocity reconstruction binned in radial bins. We see that the RMS error for the Run 1 is substantially lower than Run 2 and 3. Furthermore the RMS error on Run2 is higher than Run3. Note that only Run2 is impacted by the IHM bias. We see that the RMS error in Run3 is closer to Run2 than to Run1, thus suggesting that extra error of the IHM uncorrected run over the IHM corrected run is driven by the fact that the distance estimates are better localized when we account for the LOS density field, rather than the IHM bias. Thus, by accounting for the LOS density, we not only correct for the IHM bias, but also reduce the reconstruction error by better localization of the peculiar velocity tracers. 

